\documentclass[%
 reprint,
 amsmath,amssymb,
 aps,
]{revtex4-2}

\usepackage{graphicx}
\usepackage{dcolumn}
\usepackage{bm}
\usepackage{color} 
\usepackage{hyperref}

\usepackage{makecell}

\hypersetup{
    colorlinks = true,
    linkcolor  = blue,
    citecolor  = blue,
    urlcolor   = black
}

\makeatletter
\def\ps@plain{
   \def\@oddfoot{\hfil\thepage\hfil}%
   \def\@evenfoot{\hfil\thepage\hfil}%
   \def\@oddhead{}%
   \def\@evenhead{}%
}
\makeatother

\begin{document}

\renewcommand{\arraystretch}{1.3}  
\setlength{\tabcolsep}{5.5pt}  

\title{Spin and Orbital Edelstein effect in gated monolayer transition metal dichalcogenides\\}


\author{Tapesh Gautam}
\affiliation{Department of Physics \& Astronomy, University of Missouri, Columbia, MO 65211, USA} 

\author{S. Satpathy}
\email{satpathys@missouri.edu}
\affiliation{Department of Physics \& Astronomy, University of Missouri, Columbia, MO 65211, USA}

\begin{abstract}

The Edelstein effect consists of the non-equilibrium accumulation of magnetization in response to an applied electric field in systems with broken inversion symmetry.
While the spin Edelstein effect (SEE), originating from spin moments, is well established, its orbital counterpart, where magnetization arises from orbital moment, has only recently begun to attract attention.
In this work, we investigate the orbital Edelstein effect (OEE) in gated monolayer transition‐metal dichalcogenides (TMDs), such as MoS$_2$, by using first‐principles density‐functional calculations with both electron and hole doping.
The gate‐induced broken mirror symmetry  produces a Rashba-type chiral spin/orbital angular momentum texture, which in turn leads to the Edelstein effect in response to an applied in-plane electric field.
We find that for electron doping the Edelstein response is dominated by the orbital channel, whereas for hole doping the orbital and spin contributions are comparable.
For the case of hole doping, both OEE and SEE are strongly enhanced by a small amount of strain, due to strain-driven shifts between the $\Gamma$ and ${\rm K/K'}$ valley energies.
We derive analytical expressions for the spin and orbital Edelstein susceptibilities and evaluate their magnitudes from first‐principles.
Remarkably, the predicted OEE in gated monolayer TMDs is an order of magnitude larger than values reported in previously studied systems. Our results identify TMDs as promising platforms for studying the orbital Edelstein effect and highlight their potential applications in spintronics devices.

\end{abstract}   
\date{\today}			 		
\maketitle

\section{Introduction} 

The spin Hall effect (SHE) and the spin Edelstein effect (SEE) have been extensively studied and used for the  generation and manipulation of spin currents in spintronics devices\cite{she1, see0, see1,see2,see13,see3}.
In the SHE, a longitudinal charge current generates a spin current in the transverse direction due to spin-dependent forces which can be formulated in terms of the Berry curvatures.
In contrast, SEE is a phenomenologically similar but of entirely different physical origin, where a net spin moment is generated 
in the sample due to a non-equilibrium displacement of the Fermi surface in response to an applied electric field.
%
In addition to the spin, electrons also carry orbital moments, leading to the corresponding effects in the orbital channel, viz., the orbital Hall effect (OHE) and the orbital Edelstein effect (OEE) \cite{oee14, oee15}, where the orbital moment rather than the spin moment is responsible.

In this work, we are mainly interested in the OEE,
and its magnitude in comparison to the SEE. 
Though the OEE has been proposed quite a long time ago\cite{Levitov}, it has begun to be studied intensively only recently. 
Recent studies have suggested that the orbital Hall effect (OHE) \cite{ohe5} and orbital Edelstein effect (OEE) \cite{Levitov} can assist their spin counterparts in transport effects such as the spin-charge conversion. It has been shown that the magnitudes of orbital effects can be comparable to or even larger than their spin counterparts in some material systems, contradicting thereby earlier assumptions that their contributions were negligible  \cite{oee3,oee4,oee6}. However, despite these promising insights, the OEE predicted is generally found to be relatively small in magnitude \cite{oee4, oee8}.

In view of these recent developments, we are naturally led to study the OEE in the monolayer transition metal dichalcogenides (TMDs) such as the MoS$_2$ in the 2H structure. 
The reason for choosing these systems is that they are  particularly interesting from the orbital moment point of view due to the presence of the valley-differentiated orbital moments at the ${\rm K/K'}$ valleys in the pristine material, which leads to a large OHE\cite{ohe4, ohe5, ot2}, and they could potentially lead to a large OEE as well. 
For the Edelstein effect, one needs a Fermi surface, so that we are led to consider a doped system, with either electron or hole doping. 
Both electron and hole doping have been experimentally realized in monolayer TMDs through various approaches, including substitutional doping, surface charge transfer via molecular adsorption, electrostatic gating, intercalation techniques, etc. \cite{doping01, doping02, doping03, doping04},
so that the doped TMDs are readily available for experimental study, which provides an additional motivation for the present work.
There is one more point to consider. 
The Edelstein effect for the 2H-TMDs considered here would be zero due to their D$_{3h}$ point group, which is nongyrotropic. We therefore consider the system with the broken mirror symmetry $\sigma_h$, which makes the point group C$_{3v}$, making it gyrotropic  with a non-zero Edelstein effect. The mirror symmetry is naturally broken when the TMD is grown on a substrate, or it can be achieved by gating with an electric field applied normal to the monolayer plane.

\begin{figure} 
     \includegraphics[width=1\linewidth]{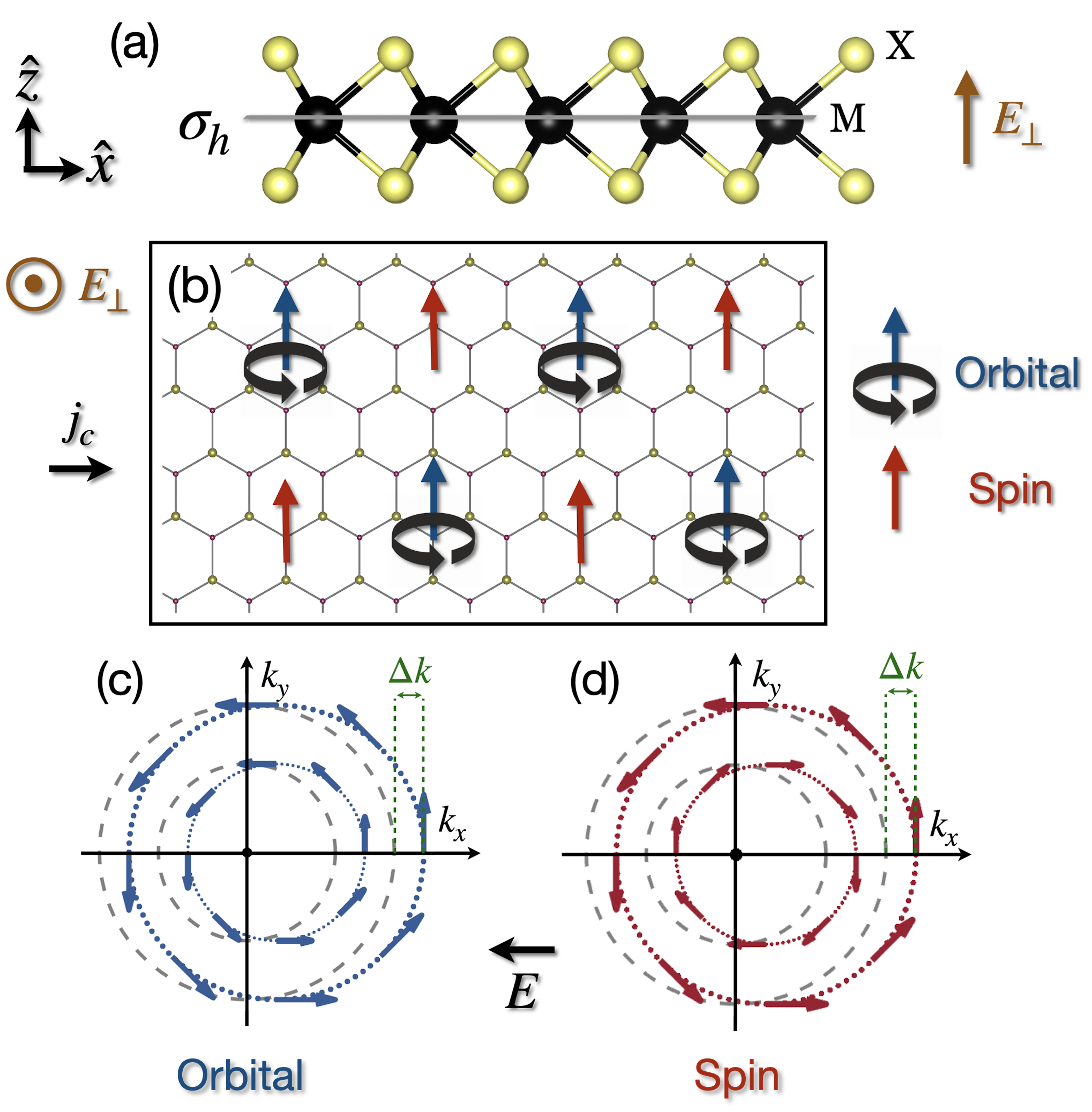}
    \caption{Orbital and spin Edelstein effects in the gated TMD MX$_2$ 
    in the 2H structure, where M = Mo, W, or Nb denotes a metal atom, and X = S, Se, or Te denotes a chalcogen atom.
    (a) Side view of the TMD monolayer with the gate field $E_\perp$ normal to the 2D plane.
    (b)  The accumulation of the  orbital/spin moments, indicated by the blue and red arrows, respectively, via the Edelstein effect under the applied in-plane charge current $j_c$. (c) and (d) The non-equilibrium shift of the Fermi surface for the hole-doped system, and the orbital and spin textures for the $\Gamma$ or ${\rm K/K'}$ valley, with  
momentum $\vec k$ measured with respect to the corresponding valley point.
     Note that the orbital angular moments of the two Fermi surfaces, split due to the spin-orbit coupling, add up (c), while the spin moments cancel (d) in such a way so as to enhance the OEE.
   }
    \label{fig:intro}
\end{figure}

The underlying physics is depicted in Figure \ref{fig:intro}. Doped MoS$_2$ serves as a representative example, where a layer of transition metal atoms (M) is sandwiched between two layers of chalcogen atoms (X). The application of an out-of-plane electric field ($E_{\perp}$) along the $z$--direction breaks the mirror symmetry relative to the monolayer plane. 
Experimentally, it is possible to apply quite large fields normal to the monolayer in the gated structures up to 4 V/nm \cite{gate01}. 
The presence of the perpendicular electric field breaks the mirror symmetry, inducing a Rashba-type chiral angular momentum texture in both the spin and orbital channels near high-symmetry points. When an electric field is applied along the monolayer plane, this chiral texture in momentum space drives the spin and orbital Edelstein effects, resulting from the non-equilibrium shift in the Fermi surface
as illustrated in the bottom part of Fig. \ref{fig:intro}. 
To study the Edelstein effect, we employed a model Hamiltonian within the five metal $d$ orbital subspace, derived from a tight-binding model that included the metal and chalcogen atoms\cite{Shanavas}, with parameters fitted using density functional theory (DFT). Our results demonstrate that in the electron-doped system, the OEE is significantly large as compared to the SEE, and dominates the Edelstein response.
For the hole-doped case, the SEE and OEE are similar in magnitude,
but both are considerably enhanced by  a small amount of strain due to the movement of the $\Gamma$ valley energy with respect to those of the ${\rm K/K'}$ valleys.

\section{Minimal electronic structure model}

 In the minimal model, we adopt an effective momentum-space Hamiltonian in the metal $d$ orbital subspace.
 This is accomplished by first obtaining a tight-binding (TB) Hamiltonian with the metal d orbitals, but including the  chalcogen orbitals as well, which is needed to enforce the symmetry of the structure, and finally removing the chalcogen orbitals, incorporating their effect via the perturbative L\"owdin downfolding\cite{Shanavas, Shanavas-downfolding}. For simplicity, it is sufficient to keep the fictitious chalcogen s orbitals, but instead if we use p or some other orbitals, the form of the Hamiltonian remains the same, since it is determined by the symmetry alone. The downfolded Hamiltonian is then fitted with the DFT band structure to obtain the Hamiltonian parameters. The resulting band structure reproduces the general features of the DFT band structure reasonably well everywhere in the Brillouin zone, including the $\Gamma$ and ${\rm K}$ valley regions, which are important for our purpose.

Thus, the procedure is to keep the metal d and chalcogen s orbitals, form the standard nearest-neighbor $ 7 \times 7$ tight-binding Hamiltonian, and downfold the chalcogen s orbitals, which results in an effective $5 \times 5$ Hamiltonian.
 In the presence of a perpendicular electric field, the same procedure is followed, except that the on-site energies $\varepsilon_s$ of the chalcogen atoms in the top and bottom layers are given an offset $\varepsilon_s \pm \delta$, which mimics the broken symmetry due to the electric field. 
The L\"owdin downforlding  procedure is outlined in the Supplementary Materials\cite{SM}.

 The resulting Hamiltonian including the SOC term is given by
\begin{equation}
    \mathcal{H} = \mathcal{H}_0 + \mathcal{H}_E  + \mathcal{H}_{soc} 
    \label{eq:hfull},
\end{equation}
where $\mathcal{H}_0$ is the Hamiltonian without the electric field, $\mathcal{H}_E$ is the additional term introduced by the electric field, and $\mathcal{H}_{soc}$ is the spin-orbit coupling.
The first part of the Hamiltonian, in the Bloch function basis
$ |\vec k\alpha \rangle = N^{-1/2} \sum_R e^{i\vec k\cdot (\vec R + \vec \tau_\alpha) } |\vec R\alpha \rangle$, where $|\vec R\alpha \rangle$ denotes the atomic orbitals, $\vec R$ is the unit cell position, $\vec \tau_\alpha$ is the orbital position in the unit cell, $\vec k$ is the Bloch momentum, and  $\alpha$ denotes the five d orbitals in the order:
$ |z^2 \rangle , | x^2 - y^2 \rangle, | xy \rangle, | xz \rangle, $ and $| yz \rangle$, is given by
 \begin{equation}
    \mathcal{H}_0  = 
\begin{pmatrix}
h_{11} &  h_{12} & h_{13} & 0 & 0 \\
h^{*}_{12} & h_{22} & h_{23} & 0 & 0 \\
h^{*}_{13} & h^{*}_{23} & h_{33} & 0 & 0 \\
0 & 0 & 0 & h_{44} & h_{45} \\
0 & 0 & 0 & h^{*}_{45} & h_{55} 
\end{pmatrix},
\label{eq:h1}
\end{equation}
where the matrix elements  are 
\begin{eqnarray}
h_{11} &=& \varepsilon_1 + t_1(2 \cos{\xi} \cos{\eta} + \cos{2 \xi}), \nonumber \\
h_{22} &=& \varepsilon_2 + 2t_2 \cos{2 \xi} + (t_2 + 3 t_3) \cos{\xi} \cos{\eta}, \nonumber \\
h_{33} &=& \varepsilon_2 + 2t_3 \cos{2 \xi} + (3 t_2 + t_3) \cos{\xi} \cos{\eta}, \nonumber \\
h_{44} &=& \varepsilon_3 + 2t_4 \cos{2 \xi} + (t_4 + 3 t_5) \cos{\xi} \cos{\eta}, \nonumber \\
h_{55} &=& \varepsilon_3 + 2t_5 \cos{2 \xi} + (3 t_4 + t_5) \cos{\xi} \cos{\eta}, \nonumber \\
h_{12} &=& t_6(\cos{2 \xi} - \cos{\xi} \cos{\eta}) + \sqrt{3} i t_7  \cos{\xi} \sin{\eta}, \nonumber \\
h_{13} &=& -\sqrt{3} t_6 \sin{\xi} \sin{\eta} + i t_7 ( \sin{2 \xi} + \sin{\xi} \cos{\eta}), \nonumber \\
h_{23} &=& \sqrt{3} (t_2 - t_3) \sin{\xi} \sin{\eta} - i t_8 \sin{\xi} (\cos{\xi} - \cos{\eta}), \nonumber \\
h_{45} &=& -\sqrt{3} (t_4 - t_5) \sin{\xi} \sin{\eta} \nonumber \\
&& \hspace{15mm}  - i t_9 \sin{\xi} (\cos{\xi} - \cos{\eta}),
\label{matrixelements1}
\end{eqnarray}
where
$(\xi,\eta) \equiv a/2 \  (k_x, \sqrt 3 k_y) $ is the scaled Bloch momentum, with $a$ being the lattice constant\cite{Note-error}.

It can be easily shown that when expanded for small momentum around the valley points ${\rm K, K'}$, the Hamiltonian assumes the familiar form\cite{xiao, ohe4} 
in the two-orbital pseudo-spin subspace $|z^2\rangle$ and $| x^2 - y^2\rangle +i \tau |xy\rangle$: 
$\mathcal{H}_0 = \vec{d} \cdot \vec{\sigma}$, which is a $2 \times 2$ matrix.
Here, $\tau =  1 $ for the ${\rm K}$ valley and $\tau = -1$ for the ${\rm K'}$ valley, $\vec \sigma$ are the Pauli matrices, and $\vec d$ is a momentum dependent constant. 
The Hamiltonian parameters $t_i$'s are combinations of the TB hopping integrals $V_{dd\sigma}$, $V_{dd\pi}$, etc. However, having obtained the form of the Hamiltonian, we obtain the parameters by directly fitting to the density-functional band structure.
The parameters are listed in Table \ref{Table1} for various compounds.
Note furthermore that due to symmetry, ${\cal H}_1$ is block diagonal, and the $| xz \rangle $ and $| yz \rangle$ orbitals do not mix with the other three orbitals. This allows us to fit the DFT bands in these two sectors separately. 
In the Supplementary Materials\cite{SM}, we compare the DFT bands with the fitted bands, which shows an overall good agreement, especially in the gap region, which is relevant to our study.

\begin{figure} [tb]
    \centering
   \includegraphics[scale = 0.3]{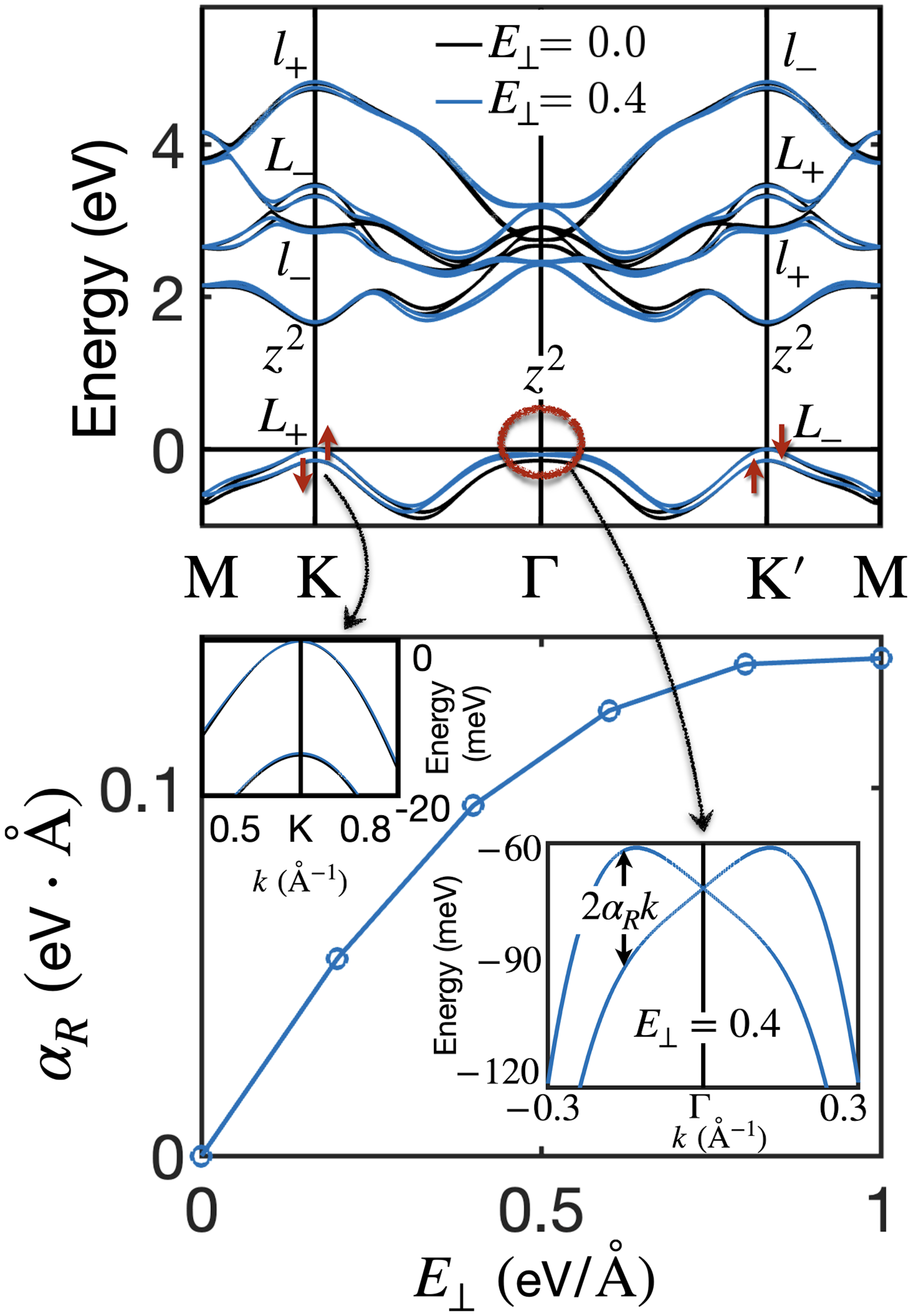}
    \caption{{\it Top,} Density-functional band structure of the MoS$_2$ monolayer (including SOC) both without and with a perpendicular electric field ($E_\perp = 0.4$ eV/\AA). {\it Bottom,} the Rashba parameter $\alpha_R$ for the splitting at $\Gamma$ as a function of $E_\perp$. {\it Bottom right inset} shows an enlarged version of 
    the Rashba-split bands at $\Gamma$, while the {\it left inset} shows very little change in the bands at the ${\rm K}$ valley, both insets being with the electric field present. 
    }
    \label{fig:bands}
\end{figure}

\begin{table*}
 \caption{Tight-binding Hamiltonian parameters obtained by fitting to the DFT bands for several TMDs. The lattice constant is $a$ (in \AA) and $\lambda$ is the SOC parameter in eV. All energies are in units of  eV. The electric field parameters $\gamma_1$ and $\gamma_2$ are listed for $E_\perp = 0.4$ eV/\AA. Both being  proportional to the electric field for small $E_\perp$, they can be calculated for other fields as well.} 
    \centering
        \begin{tabular}{c c c|c c c c c c c c|c c c c|c c}
         \hline
         \hline 
        \multicolumn{3}{c|}{} & \multicolumn{8}{c|}{Lower three bands } & \multicolumn{4}{c|}{Upper two bands } & \multicolumn{2}{c}{$E_{\perp}$ induced} \\
         \multicolumn{3}{c|}{} & \multicolumn{8}{c|}{$(|x^2-y^2\rangle, |xy\rangle, |z^2\rangle)$} & \multicolumn{4}{c|}{ $(|xz\rangle, |yz\rangle)$} & \multicolumn{2}{c}{parameters} \\
         \hline
         Material & a & $\lambda$ & $\varepsilon_1$ & $\varepsilon_2$ & $t_1$ & $t_2$ & $t_3$ & $t_6$ & $t_7$ & $t_8$ & $\varepsilon_3$ & $t_4$ & $t_5$ & $t_9$ &$\gamma_1$ & $\gamma_2$\\
         \hline
         MoS$_2$ & 3.18 &  0.075 & 1.12 & 2.15 & -0.40 & 0.05 & 0.22 & 0.96 & -0.99 & -1.34 & 3.74 & -0.20 & -0.11 & 0.95 &  0.25 &  
         {0.06} \\
         MoSe$_2$ & 3.30 & 0.093 & 0.88 & 2.03 & -0.42 & 0.12 & 0.22 & 0.85 & -0.89 & -1.16 & 3.50 & -0.05 & -0.16 & 1.05 & 0.64 &  {0.05}  \\
         MoTe$_2$ & 3.53 & 0.109 & 0.57 & 1.95 & -0.37 & 0.25 & 0.22 & 0.70 & -0.73 & -0.96 & 3.35 & 0.01 & -0.08 & 1.29 & 1.04 & {0.05}  \\
         WS$_2$ & 3.18 & 0.215 &1.20 & 2.33 & -0.42 & -0.06 & 0.29 & 1.12 & -1.00 & -1.53 & 3.99 & -0.26 & -0.11 & 0.96 & 0.17 & {0.12} \\
         WSe$_2$ & 3.32 &  0.228 & 0.92 & 2.16 & -0.41 & 0.04 & 0.26 & 0.93 & -0.96 & -1.32 & 3.73 & -0.06 & -0.22 & 1.03 &  0.32 &  {0.11}\\
         WTe$_2$ & 3.56 & 0.244 & 0.54 & 2.04 & -0.34 & 0.20 & 0.23 & 0.78 & -0.78 & -1.08 & 3.16 & 0.05 & -0.08 & 0.84 & 0.52 &  {0.03}  \\
         NbS$_2$ & 3.36 &  0.058 & 1.50 & 2.41 & -0.24 & -0.01 & 0.20 & 0.97 & -0.83 & -1.37 & 3.88 & -0.12 & -0.26 & 0.55 & 0.10 & {0.08} \\
         NbSe$_2$ & 3.46 & 0.081 &1.29 & 2.19 & -0.32 & 0.03 & 0.21 & 0.93 & -0.73 & -1.16 & 3.57 & -0.07 & -0.21 & 0.72 &  0.30 &  {0.08} \\  
         \hline
         \hline
    \end{tabular}
    \label{Table1}
\end{table*}

Similarly, the electric field part of the Hamiltonian is given by
\begin{equation}
    \mathcal{H}_{E} = 
\begin{pmatrix}
0 & 0 & 0 &  h_{14} &  h_{15}\\
0 & 0 & 0 &  h_{24} &  h_{25}\\
0 & 0 & 0 &  h_{34} &  h_{35}\\
h^{*}_{14} & h^{*}_{24} & h^{*}_{34} & 0 & 0 \\
h^{*}_{15} & h^{*}_{25} & h^{*}_{35} & 0 & 0
\end{pmatrix},
\label{eq:h2}
\end{equation}
where the matrix elements are 
\begin{align}
h_{14} &= \sqrt{3} i \gamma_1 ( e^{i \eta} \sin{\xi} + 2 \cos{\xi} \sin{\xi} ),\nonumber \\  
h_{15} &= \gamma_1 ( \cos{2 \xi} - \cos{\xi} \cos{\eta} + 3i \cos{\xi} \sin{\eta} ),\nonumber \\
h_{24} &= 2 \sqrt{3} i \gamma_2 ( - e^{i \eta} \sin{\xi} + \cos{\xi} \sin{\xi} ),\nonumber \\
h_{25} &= \gamma_2 (3 +  \cos{2 \xi} - 4 \cos{\xi} \cos{\eta} ),\nonumber \\
h_{34} &= 6 \gamma_2 \sin^2{\xi},\nonumber \\
h_{35} &= 2 \sqrt{3} i \gamma_2 ( e^{-i \eta} \sin{\xi} - \cos{\xi} \sin{\xi} ).
\label{matrixelements2}
\end{align}
As seen from the above, the electric field introduces two new parameters $\gamma_1$ and $\gamma_2$, which mixes the two d orbital sectors via the off-diagonal blocks.
This is the reason why the minimal basis consists of all five d orbitals, when the electric field effect is to be studied. The $ (|xz \rangle, |yz \rangle)$ sector cannot be further downfolded, which would have resulted in a simpler $3 \times 3$ Hamiltonian, because the on-site energies are not very different, which is needed for the 
L\"owdin downfolding. 
The final term is the SOC term   
%
\begin{align}
    &\mathcal{H}_{soc} = \lambda \ \mathbf{L} \cdot \mathbf{S} =   \frac{\lambda}{2} \times \nonumber \\ 
    &\left(
\begin{array}{ccccc|ccccc}
0 & 0 &0 &0  & 0 & 0 & 0 & 0 &  - \sqrt{3} &  i \sqrt{3}\\
0 & 0 & -2 i & 0  &  0 & 0 & 0 & 0 &  1 &  i \\
0  & 2 i & 0 & 0  &  0 & 0 & 0 & 0 &  -i &  1 \\
0  & 0 & 0 & 0  &  -i & \sqrt{3} & -1 & i &  0 &  0 \\
0  & 0 & 0  &  i & 0 & -i \sqrt{3} & -i & -1 &  0 &  0 \\
\hline 
0  & 0 & 0  & \sqrt{3}&  i \sqrt{3} & 0 & 0 &  0 &  0 &  0 \\
0  & 0 & 0  & -1 &  i  & 0 & 0 &  2 i &  0 &  0 \\
0  & 0 & 0  & -i &  -1  & 0 &  -2 i & 0 &  0 &  0 \\
- \sqrt{3} & 1 & i & 0  &  0 & 0 & 0 & 0 &  0 &  i \\
- i \sqrt{3}  & - i & 1 & 0  &  0 & 0 & 0 & 0 &  -i &  0
\end{array}
\right).
\label{eq:SOC}
\end{align}

All density functional theory (DFT) calculations were carried out using the projector augmented-wave (PAW) method\cite{Blochl}, as implemented in the Vienna Ab-initio Simulation Package (VASP) \cite{Kresse}, and employing the generalized gradient approximation (GGA) for the exchange-correlation functional.
A typical DFT band structure for the TMDs, both with and without a gate field $E_\perp$ is shown in Fig. \ref{fig:bands}, taking  MoS$_2$ as an example. 
For the TMDs, it is convenient to classify the orbital characters in terms of their angular momentum, so that in Fig. \ref{fig:bands} as well as throughout the paper, we have used the nomenclature: 
$L_\pm \equiv | x^2-y^2\rangle \pm i |xy\rangle, \  l_\pm \equiv |xz\rangle \pm i |yz \rangle, $ and $L_0 \equiv |z^2\rangle$.
The density-functional bands were fitted with the TB model, Eq. (\ref{eq:hfull}), to obtain
the Hamiltonian parameters which are listed in Table \ref{Table1} for a number of compounds. 

{\it Electric field parameters $\gamma_1$ and $\gamma_2$}.
We have extracted the electric field parameters by considering their effects on the bands at the $\Gamma$ and the ${\rm K, K'}$ valleys, which are relevant for both the electron and the hole doped cases.
We focus on the valence bands. 
At the $\Gamma$ valley, the bands are spin degenerate without the electric field,  and  the electric field causes a strong momentum-dependent Rashba splitting as seen from Fig. \ref{fig:bands}. The controlling parameter for this splitting is $\gamma_1$, the magnitude of which may therefore  be obtained from the strength of the splitting. 
Since the valence-band top at $\Gamma$ has the $|z^2\rangle$ character, we  downfold the $10 \times 10$ Hamiltonian (\ref{eq:hfull}) to the  $(|z^2 \uparrow \rangle, |z^2 \downarrow \rangle)$ sector,  which leads to the $2 \times 2$ Rashba Hamiltonian
of the form
$\alpha_R \ (k_y \sigma_x - k_x \sigma_y)$, where    
\begin{equation}
 \alpha_R = \frac{9 a \gamma_1 \lambda /2 }  {  \varepsilon_1 - \varepsilon_3+3(t_1-t_4-t_5)  + \lambda/2 }.
 \label{eq:alphaR-Gamma}
 \end{equation}
Here, only the parameter $\gamma_1$ depends on the gate field, and the remaining parameters are intrinsic to the material.
For small electric fields,  $\gamma_1$ and $\gamma_2$ are expected to vary linearly with $E_\perp$, which we have verified by fitting with the band structure under different gate fields.
This means that the Rashba splitting at the $\Gamma$ valley  $\Delta \varepsilon = 2 \alpha_R |k| $ increases with $E_\perp$, since $\alpha_R \propto \gamma_1 \propto E_\perp$. This is seen from the bottom part of Fig. (\ref{fig:bands}).

For the calculation of $\gamma_2$, we note that in contrast to the $\Gamma$ valley, valence bands at the ${\rm K/K'}$ valleys are already spin split  even without the perpendicular electric field, so that there is no linear Rashba splitting in the band structure. In view of this, we have found it convenient to compute $\gamma_2$ from the band curvature (effective mass) of the valence bands at the ${\rm K, K'}$ valleys in the absence of the SOC. 
Table \ref{Table1} lists the electric field parameters for $E_\perp = 0.4$ eV/\AA, but they can be extracted for other electric fields up to about  $E_\perp \approx 0.5$ eV/\AA, a range in which  the linear dependence is valid as seen from Fig. \ref{fig:bands}, bottom panel.

\begin{table}
\label{tab:Ising}
 \caption{The Ising and Rashba parameters, $\beta$ and $\alpha_R$, of the effective Hamiltonian Eq.(\ref{eq:Ising-Rashba}) for the valence and conduction band edges for the gate field $E_\perp = 0.4~\mathrm{eV/\AA}$. 
 %
$\beta$ is  independent of $E_\perp$ to the linear order, while 
$\alpha_R \propto E_\perp$ can be computed from the linear scaling  for other gate fields.
 Units are: eV for $\beta$ and eV$\cdot$\AA\ for $\alpha_R$, and there is no Ising term for the valence $\Gamma$ band ($\beta = 0$).} 

\begin{center}
\begin{tabular}{ c | c  c | c c | c c}
\hline
\hline 

\multicolumn{1}{c|}{Material} & \multicolumn{2}{c|}{Conduction band} & \multicolumn{4}{c}{Valance band}\\ 
\cline{2-7}
\multicolumn{1}{c|}{} & \multicolumn{2}{c|}{${\rm K/K'}$} & \multicolumn{2}{c|}{${\rm K/K'}$} & \multicolumn{2}{c}{$\Gamma$}\\ 
\cline{4-7}
\multicolumn{1}{c|}{} & \multicolumn{1}{c}{$ \beta $} & \multicolumn{1}{c|}{$ \alpha_R $} & \multicolumn{1}{c}{$ \beta $} & \multicolumn{1}{c|}{$ \alpha_R $} & \multicolumn{1}{c}{$ \beta $} & \multicolumn{1}{c}{$ \alpha_R $} \\

\hline

 MoS$_2$ &  {0.002}  & 0.2 &  {0.075} & - 0.008   & 0 & {-0.24}\\  
 MoSe$_2$ &  {0.011}  & 1.6 &  {0.093}   & - 0.011  & 0 & {-0.68}\\ 
 MoTe$_2$ &  {0.017}  & 2.7 &  {0.109} & - 0.013  & 0 & {-1.27}\\ 
 WS$_2$ &  {0.015} & 0.8 &  {0.215} & - 0.053 & 0 & {-0.46} \\
 WSe$_2$ &  {0.017}  & 1.8 &  {0.228} & - 0.062  & 0 & {-0.90}\\
 WTe$_2$ &  {0.021}  & 5.3 &  {0.244} & - 0.022  & 0 & {-1.45}\\
 NbS$_2$ &  {0.002}  & 0.2 &  {0.058} & - 0.012  & 0 & {-0.12}\\
 NbSe$_2$ &  {0.011}  & 0.9 &  {0.081} & - 0.018  & 0 & {-0.38}\\
 \hline
 \hline
\end{tabular}
\end{center}

\end{table}

{\it Effective Hamiltonian at the valence and conduction edges} -- 
The Hamiltonian in the neighborhood of the valence and conduction band edges can be written as a combination of the Ising term and a Rashba term, which is sometimes used in the literature for the gated TMDs\cite{oee11}. This can be obtained by a small-momentum expansion of the Hamiltonian Eq. (\ref{eq:hfull}) including the SOC term and then keeping the appropriate orbital state ($L_+$, $L_-$, or $z^2$) for the band edge  and L\"owdin downfolding the remaining portion of the Hamiltonian to form a $2 \times 2$ Hamiltonian in the spin space. The result is \cite{Shanavas, oee11}
\begin{equation}
    \mathcal{H}_{\rm eff} = \frac{\hbar^2 k^2} {2 m^*} + 
\tau\beta ~ \sigma_{z} +  \alpha_R ~ (k_y  \sigma_{x} - k_x \sigma_{y}),
\label{eq:Ising-Rashba}
\end{equation}
which is valid for the valence and conduction valleys at ${\rm K/K'}$ as well as the valence valley at $\Gamma$, and $m^*$ is the effective band mass. 
The second term is the Ising term, while the third term is the Rashba SOC. Both terms originate from the SOC, while the Rashba term requires a non-zero gate field as well.   In the Ising term, $\tau$ is the valley index for ${\rm K/K'}$, while it is absent for the $\Gamma$ point ($\beta = 0$).
Both the Ising parameter $\beta$ as well as the Rashba SOC strength $\alpha_R $ can be expressed in terms of the original Hamiltonian parameters listed in Table \ref{Table1}.
From these calculations, we find that to the linear order,  $\beta$ is independent of the gate field, while the Rashba term $\alpha_R \propto E_\perp$. The numerical values of  $\beta$ and $\alpha_R$ are listed in Table \ref{tab:Ising}. 
 We have computed $\beta$ directly from the spin splitting in the DFT bands with zero gate field (since it is independent of $E_\perp$ to the linear order), while  $\alpha_R $ was computed from its expression in terms of the parameters given in Table \ref{Table1} for $E_\perp = 0.4~\mathrm{eV/\AA}$. For instance, the expression for $\alpha_R$ for the $\Gamma$ valley was already given in Eq. (\ref{eq:alphaR-Gamma}). Similarly, expressions for other valleys may be obtained,
 which are listed in the Supplementary Materials\cite{SM}.

It can be shown from the downfolding procedure outlined above that for the ${\rm K/K'}$ valleys,  while the Ising term is simply $\beta = \lambda$ for the valence bands,  $\beta \propto \lambda^2$ for the conduction bands, resulting in a very small Ising splitting for the conduction ${\rm K/K'}$ valley. The small splitting is well known from earlier DFT calculations\cite{Pratik-TMD-strain}, which comes from of a perturbative mixture of the 
$|z^2 \rangle$ orbital with the $l_\pm$ angular momentum states, as indicated from the SOC matrix, Eq. (\ref{eq:SOC}).  For very small doping, either the band edge Hamiltonian Eq. (\ref{eq:Ising-Rashba}) or the full Hamiltonian Eq. (\ref{eq:hfull}) may be used.

\section{Orbital/spin texture and the Edelstein Effect}

The monolayer TMD in the 2H crystal structure has the $D_{3h}$ point group symmetry, which contains the mirror symmetry $\sigma_h$ with respect to the horizontal plane. The $\sigma_h$ symmetry  makes  $\langle L_x \rangle = \langle L_y \rangle = 0$ at all points in the Brillouin zone, while, due to the lack of inversion symmetry, $\langle L_z \rangle$ is non-zero. Indeed, as is well known, $\langle L_z \rangle$  has large values at the  ${\rm K/K'}$ valleys. 
A gate field, or alternatively a substrate, breaks the $\sigma_h$ symmetry
resulting in the $C_{3v}$ point group, and this makes all three components $\langle L_x \rangle, \langle L_y \rangle,$ and $\langle L_z \rangle \ne 0$
at every point in the BZ. The same is true for the spin moments as well.
This leads to a net accumulation of magnetic moments with components along the plane when a charge current is applied leading to the Edelstein effect. 
A similar Edelstein effect with moments $\langle L_z \rangle / \langle S_z \rangle$  normal to the plane doesn't occur because of the cancellation effect over the Brillouin zone.
and this also follows from the symmetry of the structure, as seen from Eq. (\ref{matrix}).

The magnetoelctric response tensor for the $C_{3v}$ point group is given by\cite{mag_tensor01}
\begin{equation}
    K_{ij} = \begin{pmatrix}
    0 & K_{xy} & 0 \\
    -K_{xy} & 0 & 0 \\
    0 & 0 & 0 \\
\end{pmatrix},
\label{matrix}
\end{equation}
where $\mathcal{M}_j = K_{ij} E_i$, $\mathcal {\vec M}$ being the magnetization density (magnetic moment per unit area of the crystal) produced by the in-plane
electric field $\vec E$, where subscripts denote cartesian coordinates.
Thus an electric field along $\hat x$ produces a moment $\mathcal{M}_y = K_{xy} E_x$,
while the same electric field along $\hat y$ produces the moment $\mathcal{M}_x$, where the two of them differ by a sign, $\mathcal{M}_x = - \mathcal{M}_y$.
Thus, there is a single component of the response function $K_{xy}$, which we call
$K_{\rm orb}$ and  $K_{\rm spin}$ in the orbital and the spin channel, respectively,
except that we define these to be per unit cell area $A_0$, so that
\begin{equation}
   K_{\rm orb/spin} \equiv  A_0 K_{xy}. 
\end{equation}
Thus, $K_{\rm orb}$ ($K_{\rm spin}$) is the net orbital (spin) magnetic moment developed per unit cell area of the crystal, when one unit of the in-plane electric field is applied.

In the Edelstein effect, a net magnetic moment per unit volume, $\vec {\mathcal M}$,  develops due to the non-equilibrium displacement of the Fermi surface due to the applied in-plane electric field.  In the relaxation-time approximation,
the expression is
\begin{equation}
    \vec {\mathcal M} = \frac{1}{(2 \pi)^2} \int_{\rm BZ} d^2 k \ \vec M_j (\vec{k}) \  [ f(\varepsilon_{\vec k +a_0 \vec E}) - f(\varepsilon_{\vec{k}}) ],
    \label{eq:mag}
\end{equation} 
where $\vec{ M} (\vec k) $ is the magnetization in the momentum space due to the spin moment $\vec S$ or the orbital moment $\vec L$, leading to the SEE and OEE, respectively. 
In Eq. (\ref{eq:mag}), $f (\varepsilon_{\vec k})$ is the Fermi function, $a_0  = -e \tau  / \hbar $, $\tau$ is the relaxation time  (typically $\approx 12$ ps\cite{rt2} for the TMDs), 
$- e < 0$ is the charge of the electron, and $\vec E$ is the applied electric field in the planar direction.
A Taylor series expansion of Eq. (\ref{eq:mag}) for small electric fields leads to the  response function 
\begin {equation}
   K_{ij} = \frac{e \tau }{(2 \pi)^2} \times \int_{\rm BZ} d^2 k \ v_i (\vec {k}) \  M_j(\vec {k} ) ( - \frac{\partial f_0}{\partial \varepsilon} )
     _{\varepsilon_{\vec{k}}},
\label{eq:resfn}
\end {equation}
where $\vec{v} (\vec{k}) = \hbar^{-1} \vec{\nabla} \varepsilon_{\vec k} $ is the electron velocity. For $T=0$, Eq. (\ref{eq:resfn}) reduces to an integration over the equilibrium Fermi surface.

The orbital and spin moments  $\vec M (\vec k)$   
can be computed using standard theory\cite{Niu_2005, xiao_rev} from the band energies $\varepsilon (\vec k) $ and the wave functions $u (\vec k) $ 
  \begin{eqnarray} 
  \label{mom} 
  \vec M (\vec k) 
  &=&  \frac{e}{2\hbar} {\rm Im} \  \langle \vec \nabla_k u(\vec k)| \times [{\cal H}(\vec k) -\varepsilon (\vec k)] | \vec \nabla_k u(\vec k)\rangle \nonumber \\   
  &+&  \frac{e}{\hbar} {\rm Im}\  \langle \vec \nabla_k u(\vec k) |  \times [\varepsilon (\vec k) -E_F]  | \vec \nabla_k u(\vec k)\rangle  \nonumber \\ 
 &-&  \langle u(\vec k) | {\vec m_s} |u(\vec k) \rangle.
\end{eqnarray}
The first two lines are the orbital moment contributions, while the third term is the spin moment, with  
 $\vec m_s =  g_s \mu_B \vec S $   
being the spin moment operator, and $-e < 0$ is the charge of the electron.
 The first line corresponds to the intrinsic orbital moment, while the second line comes from the field dependence of the electron density of states\cite{Niu_2005}.
Furthermore, since 
the Edelstein effect is a Fermi surface property, the second line does not contribute
and this term need not be considered any further. 
The intrinsic orbital moment is  computed by adopting the atom-centered approximation (ACA) \cite{intra_inter}, where simply the expectation value of the atomic orbital angular momentum is computed.
The ACA is often used in the literature (\cite{oee4}, \cite{ot1},  \cite{kontani_aca}), which is thought to be a good approximation in solids with atom-centered angular momentum orbitals such as the $d$ orbitals here. 
It has been shown\cite{intra_inter} to be reasonable for the TMDs for states near the band edges,
which are the relevant states in the present problem, since we consider doped systems with small dopant concentrations. 



\section{Hole doping}

\subsection{Orbital and spin textures}

\begin{figure*}
    \centering
    \includegraphics[width=0.70\linewidth]{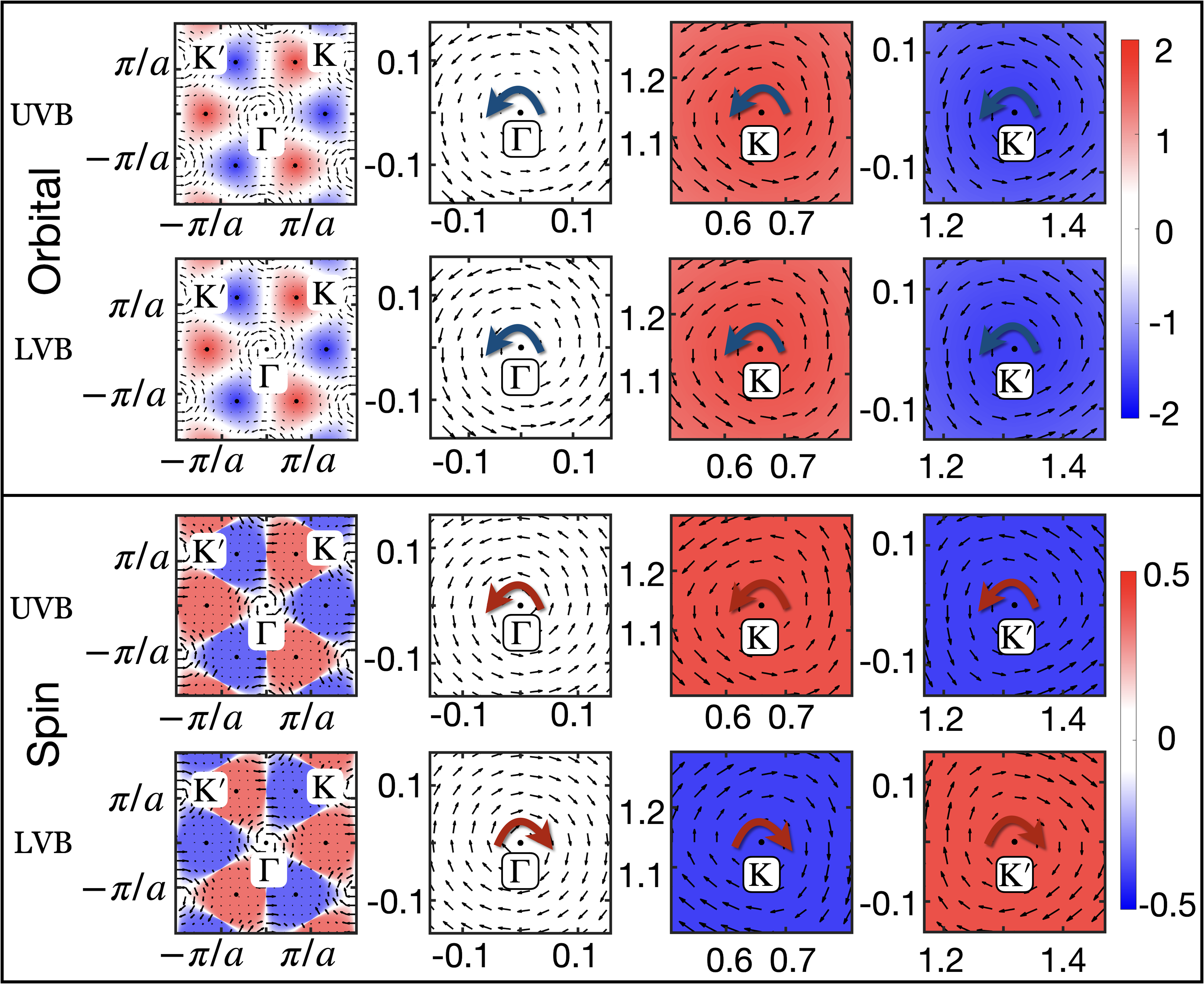}
    \caption{Orbital and spin textures of the valance bands for MoS$_2$.  The top panel shows the orbital textures for the spin-split upper and lower valence bands (UVB/LVB), respectively, at $\Gamma$ and ${\rm K/K'}$ valleys, while the lower panel shows the same for the spin texture. Thin arrows indicate the planar components $\langle L_\parallel \rangle$ or $\langle S_\parallel \rangle$, while the color coding indicates the perpendicular components $\langle L_z \rangle$ or $\langle S_z \rangle$. The bold arrows indicate the chirality of the texture. The orbital textures are the same (counterclockwise) for both UVB and LVB (top panel), while the spin textures are opposite (bottom panel). 
    This means that the Edelstein effects from the two bands add up in the orbital channel, while they have a canceling effect in the spin channel, thus enhancing the OEE.  In the Figure, the magnitude of the  gate field   $E_\perp = 0.4$ eV/\AA. }
    \label{fig:texture}
\end{figure*}

We first consider the hole-doped case.
With the effective TB Hamiltonian (\ref{eq:hfull}), we have computed the expectation values of the orbital and spin angular momentum operators, $\langle \vec L \rangle$ and $\langle \vec S \rangle$,
for the valence-band tops at the $\Gamma$ point and  the ${\rm K, K'}$ valleys in the Brillouin zone, which are shown in Figure \ref{fig:texture} for MoS$_2$.

The chiral textures are observed in both the spin and orbital channels, indicating the presence of a  Rashba-type coupling in both channels. We note that the orbital texture can come from the broken symmetry alone and it does not require the  SOC term, while the SOC term is necessary for the existence of the spin texture. When the SOC is included, the inherent orbital texture due to broken symmetry gives rise to the chiral textures in the spin channel as well.
If the SOC term is small, it can be treated perturbatively, which maintains the same orbital chirality in the two valence bands, but  the spin and orbital chiralities are opposite  for the lower valence band reflecting the gain of the SOC energy due to the $\lambda \vec L \cdot \vec S$ term.
%
This is the case for not only MoS$_2$, shown in Figure \ref{fig:texture}, but we have found the same chiralities for all TMDs we have studied.
For systems with sufficienly large SOC, which happens in some Janus materials discussed elsewhere\cite{Pratik}, the spin texture will drive the orbital texture, so that the orbital chiralities of the two bands are also opposite.

{\it Perturbative results for spin/orbital texture} -- 
The orbital/spin textures can be understood using perturbation theory for the Hamiltonian Eq. (\ref{eq:hfull}), where both the electric field and SOC are treated as small perturbations.
We will consider both  $\Gamma$ and  ${\rm K/K'}$ valleys.   
At the $\Gamma$ point, the valence top is made up of the $z^2$ orbitals as can be seen from an inspection
of the Hamiltonian Eqs. (\ref{eq:h1}) and (\ref{eq:h2}). 
For the time being, neglecting the SOC term and 
using the second-order perturbation theory, the wave function with momentum $\vec k$ in the vicinity of $\Gamma$ is given by 
\begin{equation}
    | \psi (\vec k) \rangle  = |  z^2 \rangle + 
    \sum_{\alpha \ne z^2} 
    \frac{ h_{\alpha, z^2} \ |\vec k \alpha  \rangle }{h_{z^2, z^2} - h_{\alpha,\alpha}},
    \label{eq:pert1}
\end{equation}
where $h_{ij}$ are the matrix elements of  ${\cal H}_0 + {\cal H}_E$ in Eqs. (\ref{eq:h1}) and (\ref{eq:h2}). Here, $|\vec k \alpha  \rangle$ are, again, the Bloch function basis set of the $d$ orbitals.
Computing the expectation values of the angular momentum operators with $L =2$
\begin{widetext}
\begin{equation}
    L_x = \begin{pmatrix}
        0 & 0 & 0 & 0 & \sqrt{3}i \\
        0 & 0 & 0 & 0 & i \\
        0 & 0 & 0 & -i & 0 \\
        0 & 0 & i & 0 & 0 \\
        -\sqrt{3}i & -i & 0 & 0 & 0 \\
    \end{pmatrix}, \ 
   L_y = \begin{pmatrix}
        0 & 0 & 0 & -\sqrt{3}i & 0\\
        0 & 0 & 0 & i & 0 \\
        0 & 0 & 0 & 0 & i \\
        \sqrt{3}i & -i & 0 & 0 & 0 \\
        0 & 0 & -i & 0 & 0 \\
    \end{pmatrix}, \ 
   L_z = \begin{pmatrix}
        0 & 0 & 0 & 0 & 0\\
        0 & 0 & -2i & 0 & 0 \\
        0 & 2i & 0 & 0 & 0 \\
        0 & 0 & 0 & 0 & -i \\
        0 & 0 & 0 & i & 0 \\
    \end{pmatrix},
\end{equation}
\end{widetext}
and keeping the lowest order terms in momentum $\vec k$, we find the elegant result 
 \begin{equation}
    \langle \vec L_\parallel \rangle_\Gamma =  c \gamma_1 k \hat \theta,
      \label{eq:L-Gamma}
 \end{equation}
where $\theta$ is the polar angle on the 2D plane, so that  
    $ \langle L_x \rangle_\Gamma = - c\  \gamma_1 \ k \sin{\theta}$ 
      and $\langle L_y \rangle_\Gamma = c\   \gamma_1 \ k \cos{\theta}$, 
while  it can be easily shown that $ \langle L_z \rangle_\Gamma = O (k^2)$. 
In Eq. (\ref{eq:L-Gamma}),  $ c \equiv  9  a \times [\varepsilon_3 - \varepsilon_1 + 3(t_1-t_4-t_5)]^{-1} > 0 $ and $\gamma_1 \propto E$  for small electric fields, which leads to a chiral orbital texture in the counter-clockwise direction as seen in Figure \ref{fig:texture}. When the SOC term $\lambda \vec L \cdot \vec S$ is added on top of this as an additional perturbation, 
since $ \lambda > 0$, the upper (lower) band  has its spin aligned in the same (opposite) direction to the orbital texture. Thus, we have 
\begin {equation}
\langle \vec S_\parallel \rangle_\Gamma =   \frac{1}{2} \ \nu \  \hat \theta,
\label{eq:S-Gamma}
\end{equation}
where $\nu = \pm 1$ indicates the UVB and LVB, respectively. These explain the spin/orbital textures for the $\Gamma$ valley in Fig. \ref{fig:texture}.

The  textures at the  ${\rm K/K'}$ valleys can be evaluated  using the same framework. The only difference is that we have to first perform a unitary transformation of the Hamiltonian from the $(|x^2-y^2 \rangle, |xy\rangle)$ to the $(L_+, L_-) \equiv (|x^2-y^2 \rangle + i |xy \rangle, \ |x^2-y^2 \rangle - i |xy \rangle )$ basis, and then perform the perturbation theory for the $L_+$ ($L_-$) state at ${\rm K}$ (${\rm K'}$) point, which forms the valence band top. The perturbation theory, analogous to Eq. (\ref{eq:pert1}) but now applied to the $L_+ / L_-$ state, yields after straightforward algebra the result 
 \begin{equation}
    \langle \vec L_\parallel \rangle_{\rm K} =
    \langle \vec L_\parallel \rangle_{\rm K'} =  c' \gamma_2 q \hat \theta,
      \label{eq:L-K}
 \end{equation}
 while $ \langle L_z \rangle_{\rm K} = - \langle L_z \rangle_{\rm K'} = 2 + O (q^2)$. 
 Here, $\vec q = \vec k - \vec K$ or $\vec k - \vec K' $ is the momentum measured from the respective valley point and
 the constant
 $ c' =  3 \sqrt 3  a \times 
[(\varepsilon_3 - \varepsilon_2) + \frac{3}{2}(t_2+t_3-t_4-t_5) - 3 \sqrt 3 (t_8-t_9)/4 ]^{-1} $ is a positive quantity, so that the chirality of the orbital momentum in the plane is again counter-clockwise just like for the $\Gamma$ point. Note that the orbital texture increases linearly with $q$ for small momentum and its strength is controlled by the electric field parameter $\gamma_2 \propto E$, while for the $\Gamma$ valley as seen above, it was controlled by $\gamma_1$. 

Treating the SOC  as a perturbation, the diagonalization of the $ 2 \times 2$ matrix within the $(L_+ \uparrow, L_+ \downarrow)$ subspace, downfolded from the full $10 \times 10$ Hamiltonian (\ref{eq:hfull}),  leads to the spin texture 
 \begin{equation}
    \langle \vec S_\parallel \rangle_{\rm K} = 
    -  \langle \vec S_\parallel \rangle_{\rm K'} =
    \nu  c' \gamma_2  q \hat \theta,
\label{eq:S-K}
 \end{equation}
while $\langle S_z \rangle_{\rm K} = - \langle S_z \rangle_{\rm K'}  = \nu /2 + O (q^2)$,
where, again, $\nu = \pm 1$ indicates the UVB and LVB, respectively. 
Note that in all cases, the spins are aligned with $\vec L$ for the UVB and anti-aligned with it for the LVB, as may be expected from the energetics of the $\lambda \vec L \cdot \vec S$ SOC term.
These results are consistent with the spin/orbital textures seen in Fig. \ref{fig:texture}, which were calculated from the full Hamiltonian. For momentum points away from $\Gamma$ or ${\rm K/K'}$, there is deviation from these results due to warping of the band structure.

\subsection{ Edelstein susceptibility}

\begin{figure}
    \centering
    \includegraphics[scale=0.16]{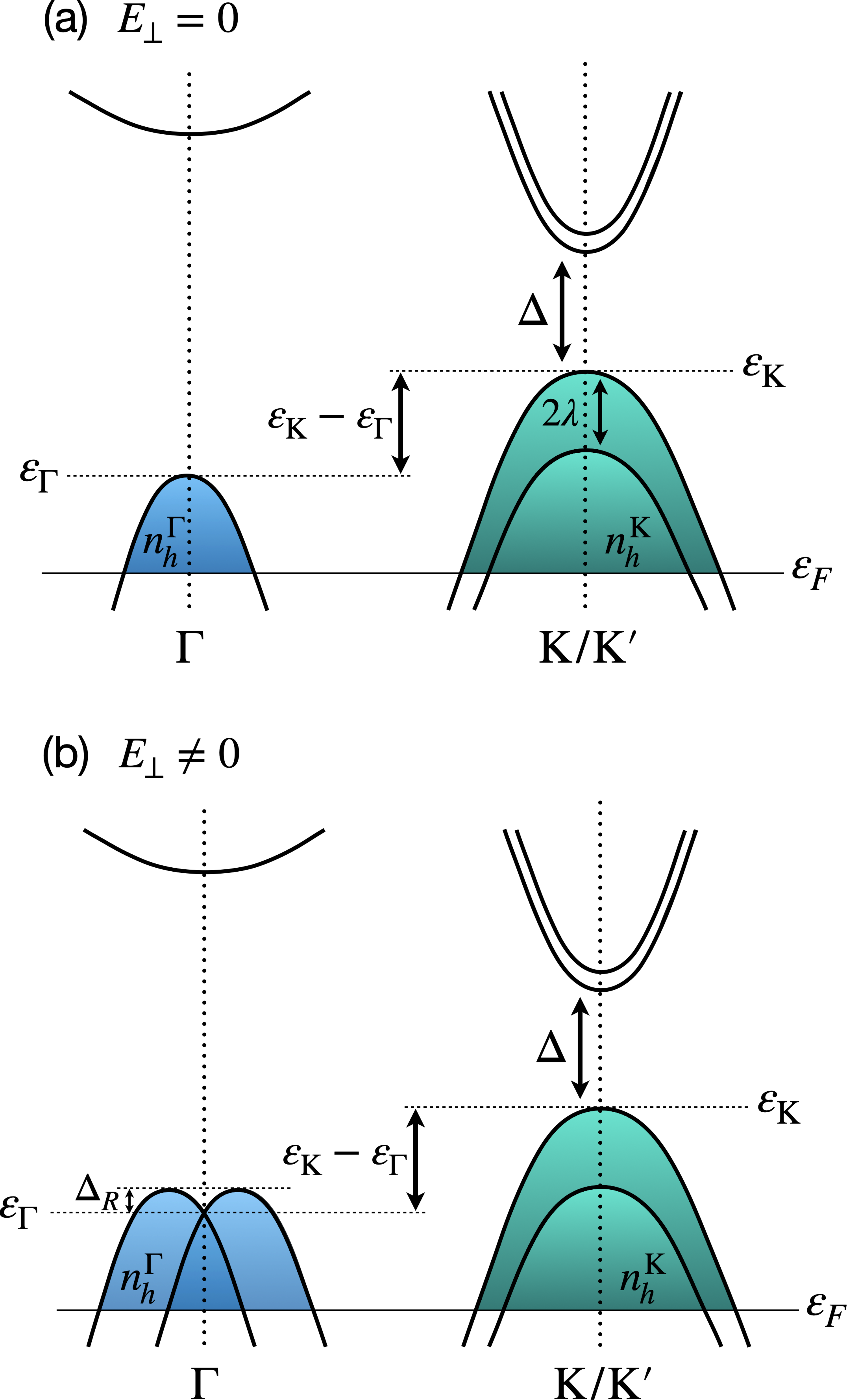}
    \caption{Schematic spin splitting for the gated TMDs, modeled by the Ising and the Rashba terms in Eq. (\ref{eq:Ising-Rashba}): (a) $E_\perp = 0$ and (b) $E_\perp \ne 0$. 
    The main effect of the gate field $E_\perp$ is to produce a Rashba splitting at the $\Gamma$ valence bands. The ${\rm K/K'}$ bands, both valence and conduction, also experience a Rashba interaction, but its effect is barely visible on account of the already-existing spin splitting. 
    Here, $\varepsilon_F$ shown is the Fermi energy for the hole doped case.
    For the electron doped case, electrons occupy the ${\rm K/K'}$ bands, with the $\Gamma$ band remaining empty owing to its much higher energy.
    A  compressive strain reduces the $\varepsilon_K - \varepsilon_\Gamma$ energy
    separation and results in a much stronger Edelstein effect in the hole doped system as discussed in the text.
}
\label{schematic-2}
\end{figure}

The Edelstein susceptibility  is easily evaluated from the analytic expressions for the magnetization, Eqs. (\ref{eq:L-Gamma}-\ref{eq:S-K}). 
We first perform the integration of Eq. (\ref{eq:resfn}) over the Fermi energy and then express the Fermi energy in terms of hole density by
using the well-known expression for the density-of-states in two dimensions.
Including both spins, it is given by the expression
$  \rho_{2D}  (\varepsilon) =  m A/ (\pi \hbar^2)$, $A$ being the area of the 2D crystal,
which is valid for parabolic bands and therefore for small dopant concentrations.
The results for the individual valleys and bands, expressed 
in units of 
$    e \tau \hbar^{-1} A_0$, are given by 
\begin{equation}
K_{\rm orb} = 
    \begin{cases}
       c \gamma_1 n_h^{\Gamma \nu} & \ \  \ \ \ \ \ \  \ \ \text{($\Gamma$ valley)}\\
     c' \gamma_2 n_h^{K \nu} & \ \ \ \ \ \text{ (K/K$^\prime$ valley)} 
    \end{cases} 
    \label{Korb}
\end{equation}
for the orbital Edelstein effect, and
\begin{equation}
K_{\rm spin} = 
    \begin{cases}
         \nu (4\pi)^{-1/2 } \sqrt  { n_h^{\Gamma \nu}} & \ \  \ \ \ \ \ \  \ \ \text{($\Gamma$ valley)}\\
     \nu \times  
    c' \gamma_2 n_h^{K \nu} & \ \ \ \ \ \text{ (K/K$^\prime$ valley)},
    \end{cases}  
    \label{Kspin}
\end{equation}
for the spin Edelstein effect. 
Notice that $K_{\rm orb}$ is the same for both bands, $\nu = \pm 1$ for the UVB and LVB, respectively,
while $K_{\rm spin}$ has the opposite sign between the two bands.
The total susceptibility is sum over the hole states in both valleys as well as both bands. 
Here, 
$n_h^{\Gamma \nu}$/ $n_h^{{\rm K} \nu}$ are the hole densities (number of holes per unit area) in the $\Gamma$ or ${\rm K}$ pocket corresponding to the specific band $\nu$. The total hole density $n_h$ is a sum of these:
$n_h = n_h^{\Gamma +} + n_h^{\Gamma -} + n_h^{\rm K +} + n_h^{\rm K-}$.
Note that the hole densities are slightly different for the UVB and LVB $\nu = \pm 1$ due to the spin splitting as seen from Fig. \ref{schematic-2}.

\begin{figure}
      \includegraphics[scale=0.25]{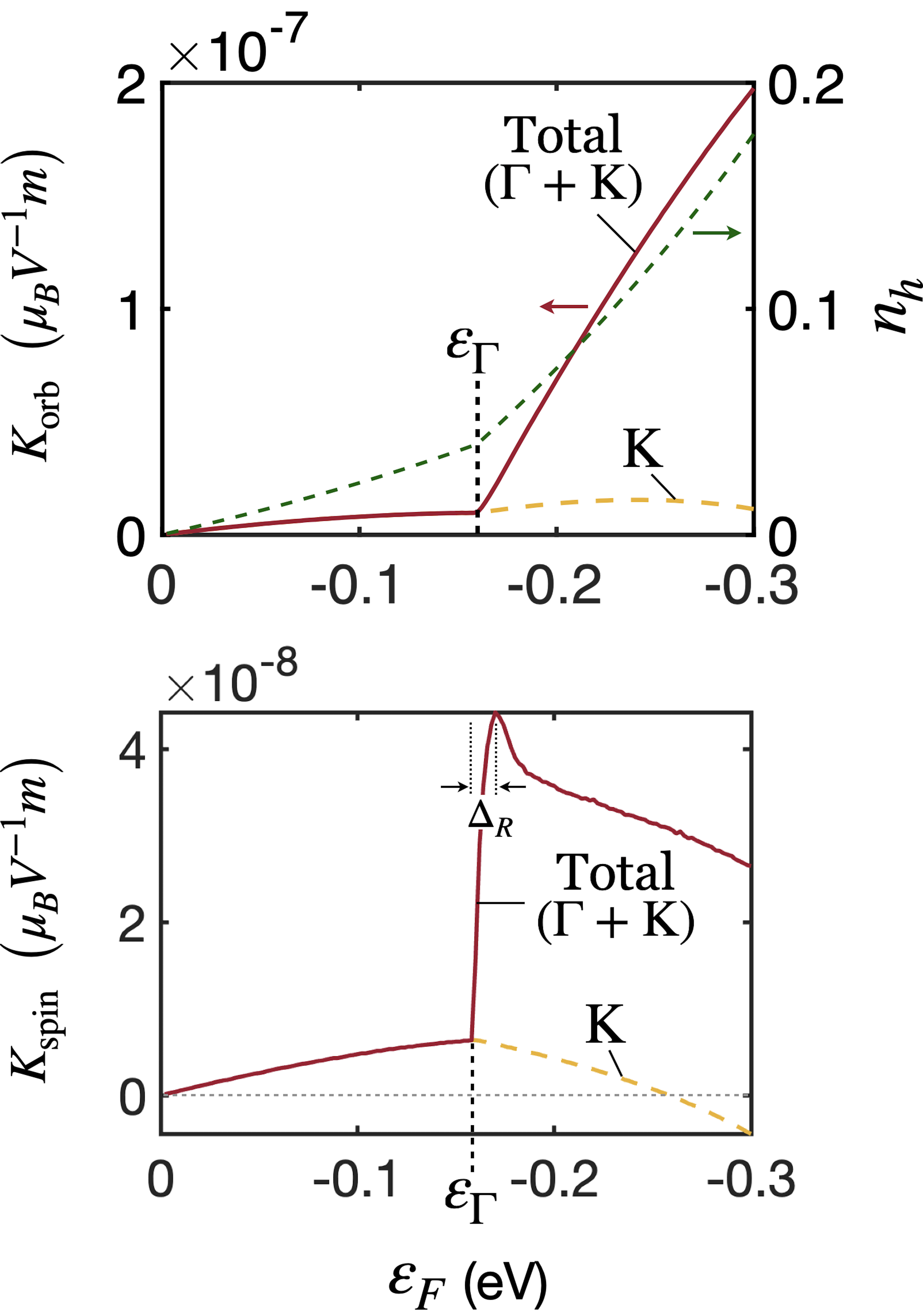}    
    \caption{Edelstein effect in the hole-doped MoS$_2$ as a function of  $\varepsilon_F$. $K_{\rm orb}$ and $K_{\rm spin}$ are the Edelstein susceptibilities,
     i.e., the magnetization $\mathcal{M}_y$ (in units of $\mu_B$) that develops   per unit cell area of the crystal, if an in-plane electric field of 1 V/m is applied along $\hat x$.
     Both the total response ($\Gamma + {\rm K/K'}$) and the partial response from the ${\rm K/K'} $ valleys are shown. Because $\gamma_1 >> \gamma_2$, contribution from the $\Gamma$ valley dominates, and therefore the response becomes much stronger once holes begin to occupy the $\Gamma$ band.
    $\Delta_R$ is the Rashba energy gain, beyond which {\it both} the Rashba bands at $\Gamma$ are occupied, and their cancellation effect kicks in, resulting in a progressive decline in  $K_{\rm spin}$ with $\varepsilon_F$ as seen in the {\it bottom} figure.
    The parameters are: $E_\perp = 0.4$ V/\AA\ and $\tau = 16$ ps (measured value for MoS$_2$\cite{rt2}), and  energy of the valence-band top is set to zero. 
    } 
    \label{fig:results}
\end{figure}

%
The linear $n_h$ dependence in Eqs. (\ref{Korb}-\ref{Kspin}) can be traced back to the linear momentum dependence of the spin/orbital textures in Eqs. (\ref{eq:L-Gamma}-\ref{eq:S-K}),
while the $\sqrt n_h$ dependence in Eq. (\ref{Kspin}))
is due to  $\langle \vec S_\parallel \rangle_\Gamma $ in Eq. (\ref{eq:S-Gamma}), which is momentum independent.
%
The susceptibility expressions Eqs. (\ref{Korb}-\ref{Kspin}) convey the essential physics of the  Edelstein effect, in particular, how they depend on  the electric field and the hole concentration, and they 
agree quite well with the DFT results presented in Fig.  \ref{fig:results}. 

Fig.  \ref{fig:results} shows the computed Edelstein response computed from Hamiltonian Eq. (\ref{eq:hfull}) as a function of the Fermi energy of the hole pocket $\varepsilon_F$ (measured with respect to the valence band top $\varepsilon_K$).
Since valence top occurs at ${\rm K/K'}$ points, as $\varepsilon_F$ is lowered, holes first occupy the ${\rm K/K'}$ pockets and once $\varepsilon_F$ falls below $\varepsilon_\Gamma$, the $\Gamma$ pocket starts filling up and begins to contribute to the Edelstein response. We note two main points from the Figure.

(i) The $\Gamma$ pocket dominates both the OEE and SEE, and once holes occupy the $\Gamma$ pocket, both the effects are dramatically enhanced. This follows from the fact that the parameter $\gamma_1 $ that determines the response at $\Gamma$ is much larger than $ \gamma_2$ (see Table \ref{Table1}), which determines the effect at ${\rm K/K'}$.
For the $\Gamma$ pocket to be occupied,  the hole concentration must exceed a critical value $ n_h > n_h^c$,
which are listed in Table \ref{Table2} for a number of materials for zero gate field. These numbers may change somewhat in the presence of a gate field $E_\perp \ne 0$.
In some cases, the hole doping needed is too high to be experimentally feasible. However, it turns out that by applying a small strain, the energy difference between $\varepsilon_\Gamma$ and $\varepsilon_K$ can be reduced or even inverted, so that even for a small hole concentration, the $\Gamma$ pocket is occupied, which therefore results in a large  OEE and SEE, as discussed in the next section.
 
(ii) The OEE from the two subbands is additive, being of the same sign, while 
the SEE from the two subbands almost cancel each other, being of the opposite sign. This is seen clearly from the Edelstein susceptibility
Eqs. (\ref{Korb}, \ref{Kspin}), where $\nu = \pm 1$ represents the two subbands. 
The fundamental reason for this is  that the orbital moments
have the same chirality for the two subbands, while the spin moments have the opposite chiralities
as seen from the spin/orbital textures, Fig. \ref{fig:texture} and also from the analytical expressions Eqs. (\ref{eq:L-Gamma}-\ref{eq:S-K}). The cancellation effect is prominently seen for the $\Gamma$ pocket in the figure for $K_{\rm spin}$. Once the $\Gamma$ pocket starts to fill ($\varepsilon_\Gamma > \varepsilon_F$), $K_{\rm orb}$ increases dramatically due to this additive effect. 
In contrast, $K_{\rm spin}$ first increases, but then starts to decrease once the lower valence band starts to be filled, and the two contributions nearly cancel. With $m^*$ denoting the band effective mass, the separation between the two Rashba-split bands is denoted by $  \Delta_R \approx  (2 \hbar^2)^{-1} m^{*2} \alpha_R^2   $ in Fig. \ref{fig:results}, beyond which the cancellation effect occurs. They don't exactly cancel because the number of holes $n_h$ is somewhat different for the two subbands due to the energy splitting, leading to a peaked structure in $K_{\rm spin}$, which gradually reduces to zero as $\varepsilon_F$ is lowered.

\subsection{Strain enhancement of the Edelstein susceptibility for the hole doped case}

Since the Edelstein effect from the $\Gamma$ valley is dominant in the hole-doped systems, the effect can be enhanced by strain.
Strain changes the 
energy separation between the $\Gamma $ and the ${\rm K/K'}$ valleys, so that the holes can occupy the $\Gamma$ pocket, which has a much larger Edelstein response as was seen from Fig. \ref{fig:results}.
We predict that for the same hole concentration, a small compressive strain leads to a significant enhancement of both the OEE and the SEE. 

\begin{figure}[bt]
    \centering
    \includegraphics[scale=0.3]{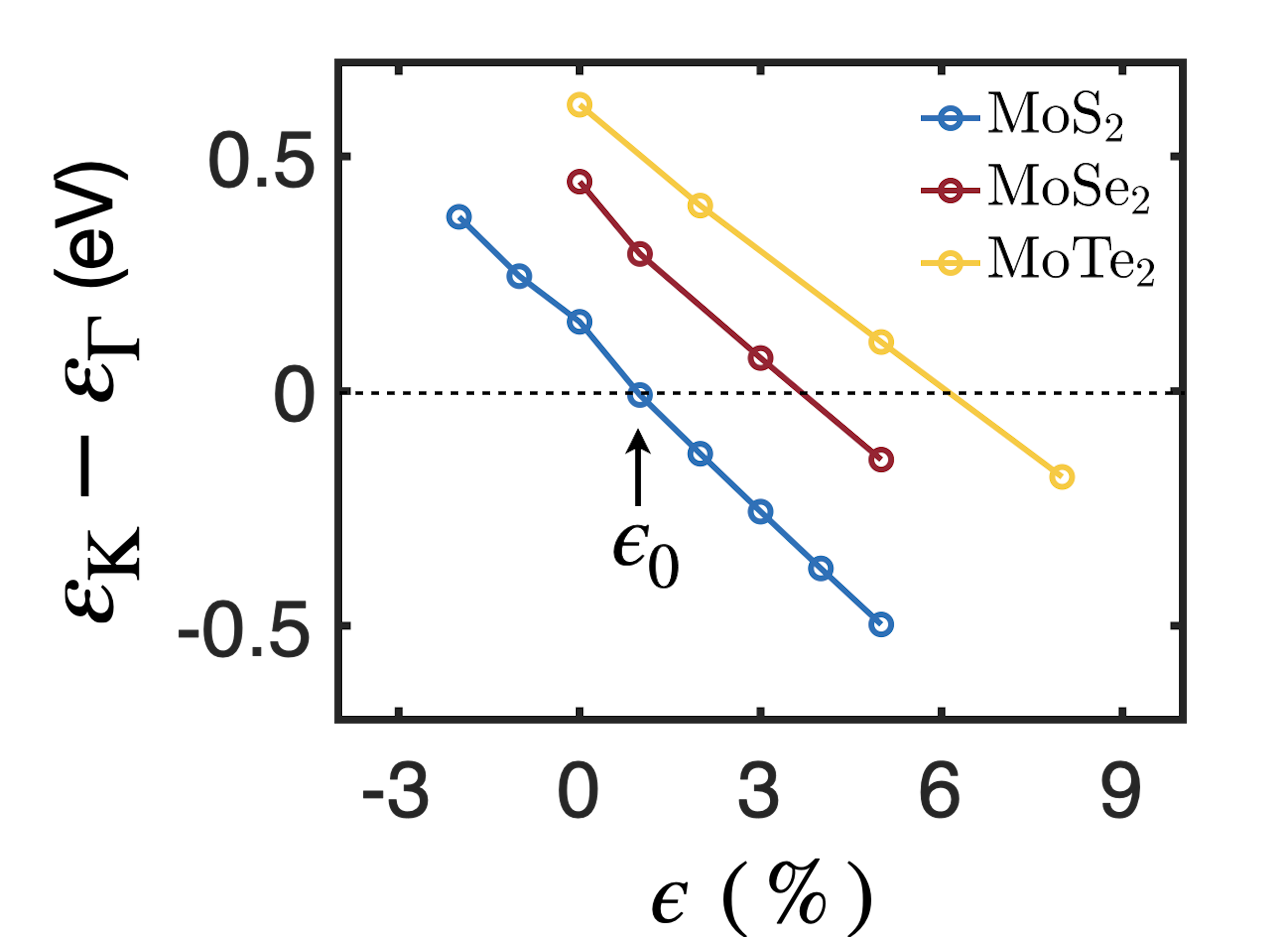}
    \caption{Valence-band valley separation $\varepsilon_K - \varepsilon_\Gamma$ as a function of uniform compressive strain $\epsilon$. Beyond a critical strain ($\epsilon > \epsilon_0$), the $\Gamma$ point forms the valence band top.
     Arrow indicates $\epsilon_0 \approx 1 \%$ for MoS$_2$.
    With increasing strain, the doped holes move to the $\Gamma$ pocket from the ${\rm K/K'}$ pockets, enhancing thereby both the OEE and the SEE.
   }
    \label{fig:E-strain}
\end{figure}

\begin{table}
 \caption{Strain effects on the OEE and SEE under uniform compressive strain $\epsilon$.  
 For zero strain, 
 $ \varepsilon_K - \varepsilon_\Gamma$ is the energy separation between the two valence band valleys, while $N^h_c \equiv n_h^{c} A_0$ is the critical hole concentration (number of holes per unit cell) needed to populate the $\Gamma$ pocket.
 $\epsilon_0$ is the strain needed to close the $\Gamma-{\rm K}$ gap.
 $K_{\rm orb}$ and $K_{\rm spin}$ are the Edelstein susceptibilities when the number of holes in the $\Gamma$ pocket is $N_h = 0.02$  per unit cell area and the
 gate field is $E_\perp = 0.4$ eV/\AA. Here, $\tau \approx 12$ ps, and $K_{\rm orb}$ and $K_{\rm spin}$ are in units of $ 10^{-8} \mu_B V^{-1}m$.
 }
    \centering
    \begin{tabular}{c | c c c | c c}
        \hline
        \hline 
         Material & $\varepsilon_K - \varepsilon_\Gamma  (eV)$ & $N^h_c$ & $\epsilon_0 (\%)$  &   $K_{\rm orb}$ & $K_{\rm spin}$  \\ \hline
         MoS$_2$ & {0.15}  & {0.03} & {1.0}   & {7.97}  & {2.30} \\
         MoSe$_2$ & {0.45}  & {0.43} &  {3.8} & {18.24} & {3.43} \\
         MoTe$_2$ & {0.61}  & {1.33} & {6.0} & {26.84} & {3.38}  \\
         WS$_2$ & {0.26}  & {0.05} & {2.2}    & {4.44} & {4.98}\\
         WSe$_2$ & {0.49}  & {0.14} & {4.8}  & {7.94} & {9.01}\\
         WTe$_2$ & {0.61}  & {0.47} & {7.3}  & {12.94} & {12.71}\\
         
         \hline
         \hline
    \end{tabular}
    \label{Table2}
\end{table}

\begin{figure}[tb]
    \centering
   \includegraphics[scale=0.26]{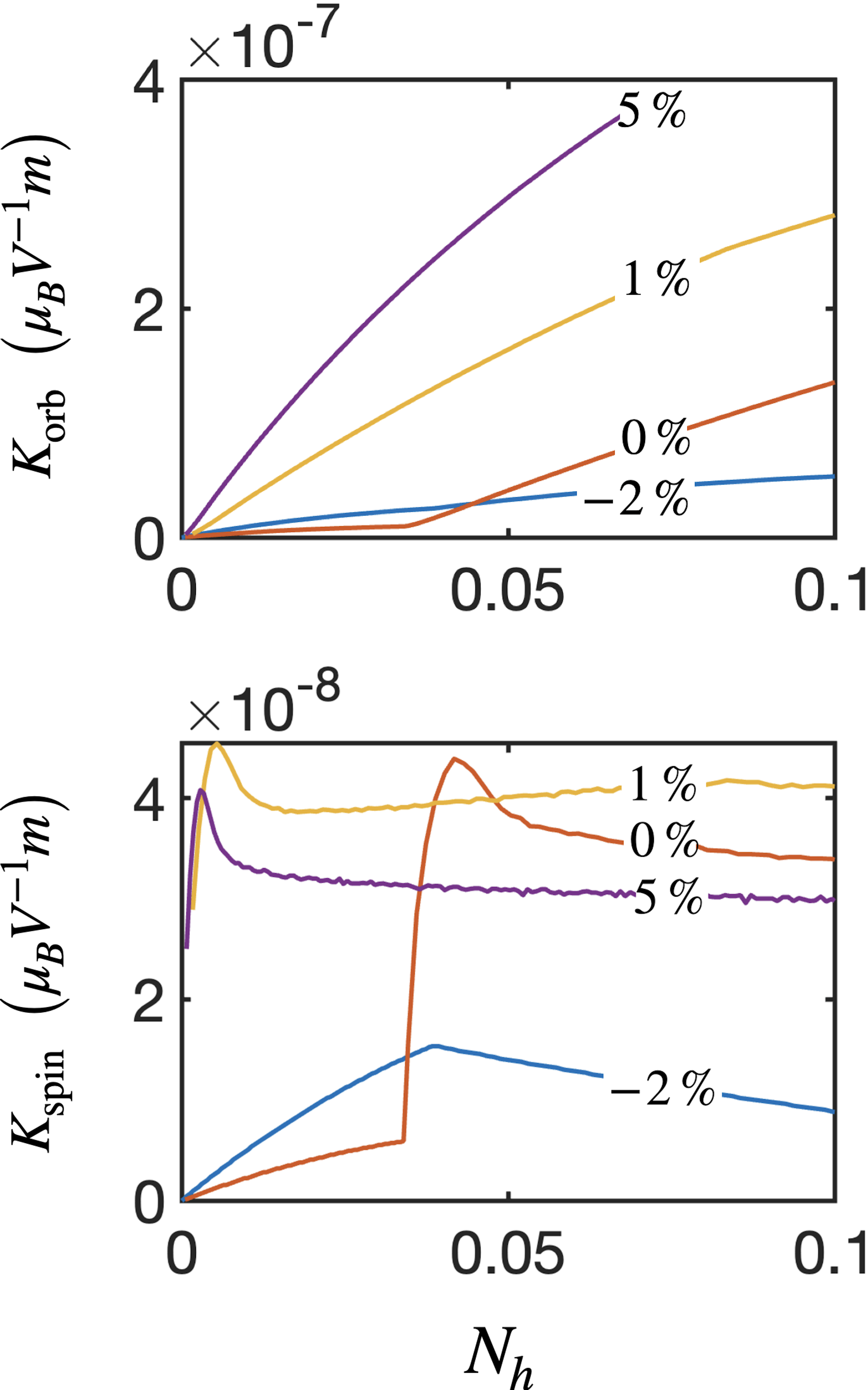}
    \caption{Orbital and spin Edelstein susceptibilities for various strains $\epsilon$  
    (listed as a percentage) as a function of the hole density $N_h \equiv n_h A_0$ 
    (number of holes per unit cell area) for gated MoS$_2$ ($E_\perp = 0.4$ eV/\AA).
    For a fixed hole density, both $K_{\rm orb}$ and 
$K_{\rm spin}$ are significantly enhanced by applying a small amount of strain.
    }
    \label{fig:strain-results}
\end{figure}

We consider the case of the uniform strain. 
Electronic structure of the TMDs under strain has been studied in great detail\cite{Pratik-TMD-strain}, and it is 
well known that  strain shifts the valence-band energies of the ${\rm K}$ and $\Gamma$ valleys. 
Fig. \ref{fig:E-strain} shows the relative positions of the two valleys $\varepsilon_K - \varepsilon_\Gamma$ as a function of strain for several TMDs,
where $\epsilon > 0 $ denotes a compressive strain.
Beyond a critical amount of compressive strain $\epsilon_0$, which are just a few percent as listed in the Table \ref{Table2}, the $\Gamma$ pocket forms the valence band top.
The strain parameter is defined simply as $\epsilon = (a_s - a)/a$, where $a_s (a)$ is the lattice constant of the  strained (unstrained) crystal.

In the unstrained system, since the valence band maximum is located at ${\rm K/K'}$, with much lower Edelstein response than the $\Gamma$ pocket, achieving a strong Edelstein effect taking advantage of the $\Gamma$ pocket requires higher doping levels. The hole concentration in the unstrained structure needed to populate the $\Gamma$ pocket, indicated by $N_c^h$ in Table \ref{Table2}, is experimentally achievable for some systems 
like MoS$_2$ ($N_c^h \approx 0.03$ per unit cell area).
We note that for the TMDs, hole doping up to as high as $ 4 \times 10^{14} {\rm cm}^{-2}$ 
($N_h \approx 0.35$ per unit cell area) has been experimentally demonstrated\cite{doping01}. However, for many TMDs, the needed $N_c^h$ is way too large. In these systems, a small strain can be applied, which can enhance the Edelstein response by an order of magnitude.
The computed Edelstein  susceptibility as a function of strain and  hole density are shown in Fig. \ref{fig:strain-results}, which shows a large enhancement with strain.

\begin{figure} 
     \includegraphics[scale=0.3]{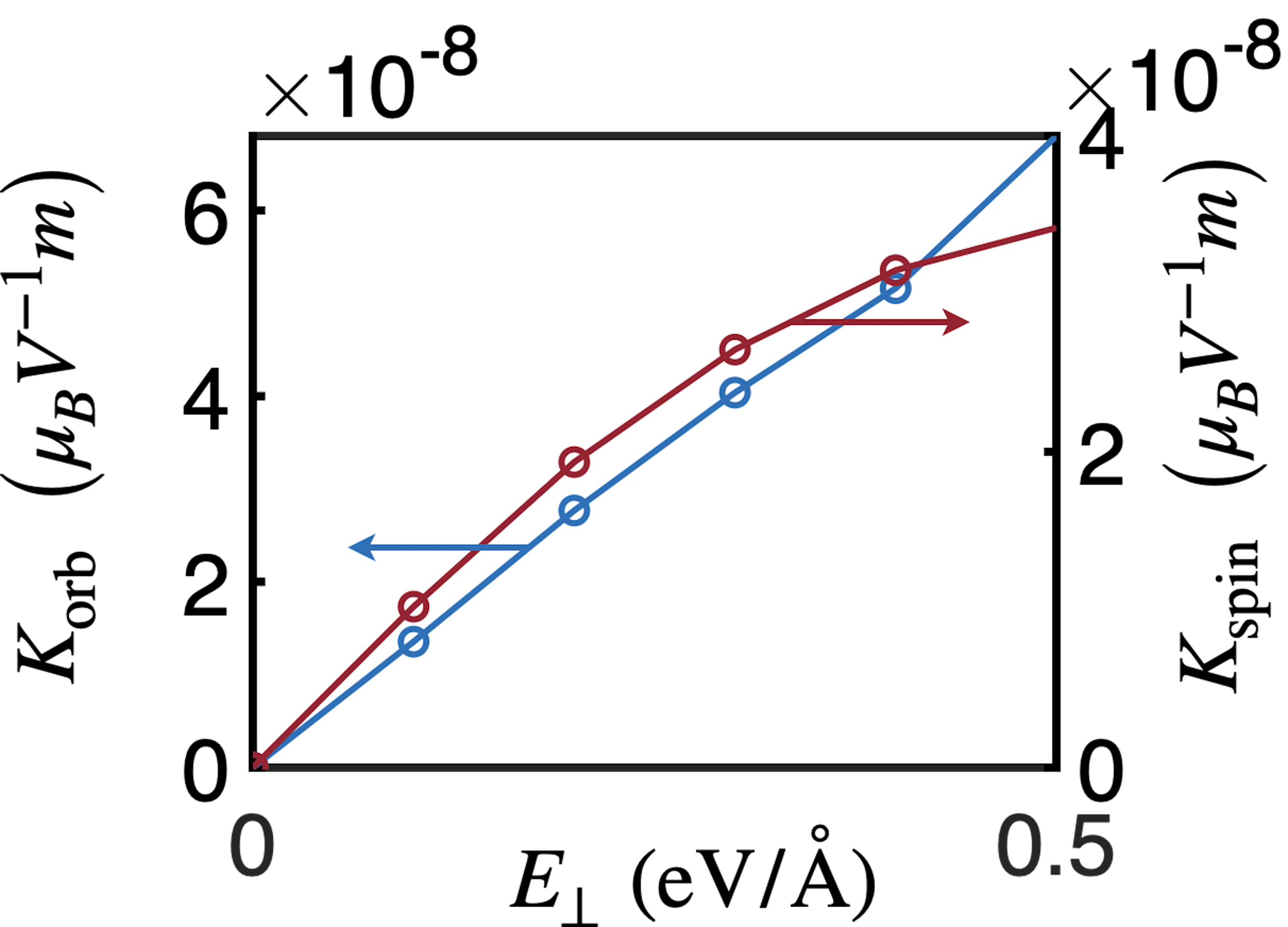} 
   \caption{Variation of the orbital and spin Edelstein response with the gate field $E_\perp$ for MoS$_2$ for hole doping with $N_h = 0.01$ holes per unit cell in the $\Gamma$ pocket.} 
   \label{fig: Gate-dependence-hole}
\end{figure}

The strength of the Edelstein response for different TMDs differ depending   on the material parameters. In order to quantify this, 
in Table \ref{Table2}, we have listed 
 the Edelstein susceptibility for various materials for a fixed number of hole concentration $N_h = 0.02$ in the $\Gamma$ pocket, which may be achieved with a small number of holes in the strained structure, or with a larger number of holes in the unstrained structure so as to populate the $\Gamma$ pocket. According to these results, 
 $K_{\rm orb}$ is large in MoTe$_2$, however a strain in excess of $\epsilon > 6 \%$ is needed to populate the $\Gamma$ pocket, while for MoS$_2$, though $K_{\rm orb}$ is smaller,
 the strain needed to achieve this is only about 1\%. These considerations help in the choice of materials to study the Edelstein effect. 
 Finally, Fig. \ref{fig: Gate-dependence-hole} shows the dependence of the Edelstein susceptibility on the gate field, the dependence being linear for small fileds as anticipated. 
These results demonstrate the potential for strain engineering of the Edelstein effect, which may be  of potential value in spintronics applications.

\section{Electron doping}

\begin{figure}[tb]
     \includegraphics[scale=0.3]{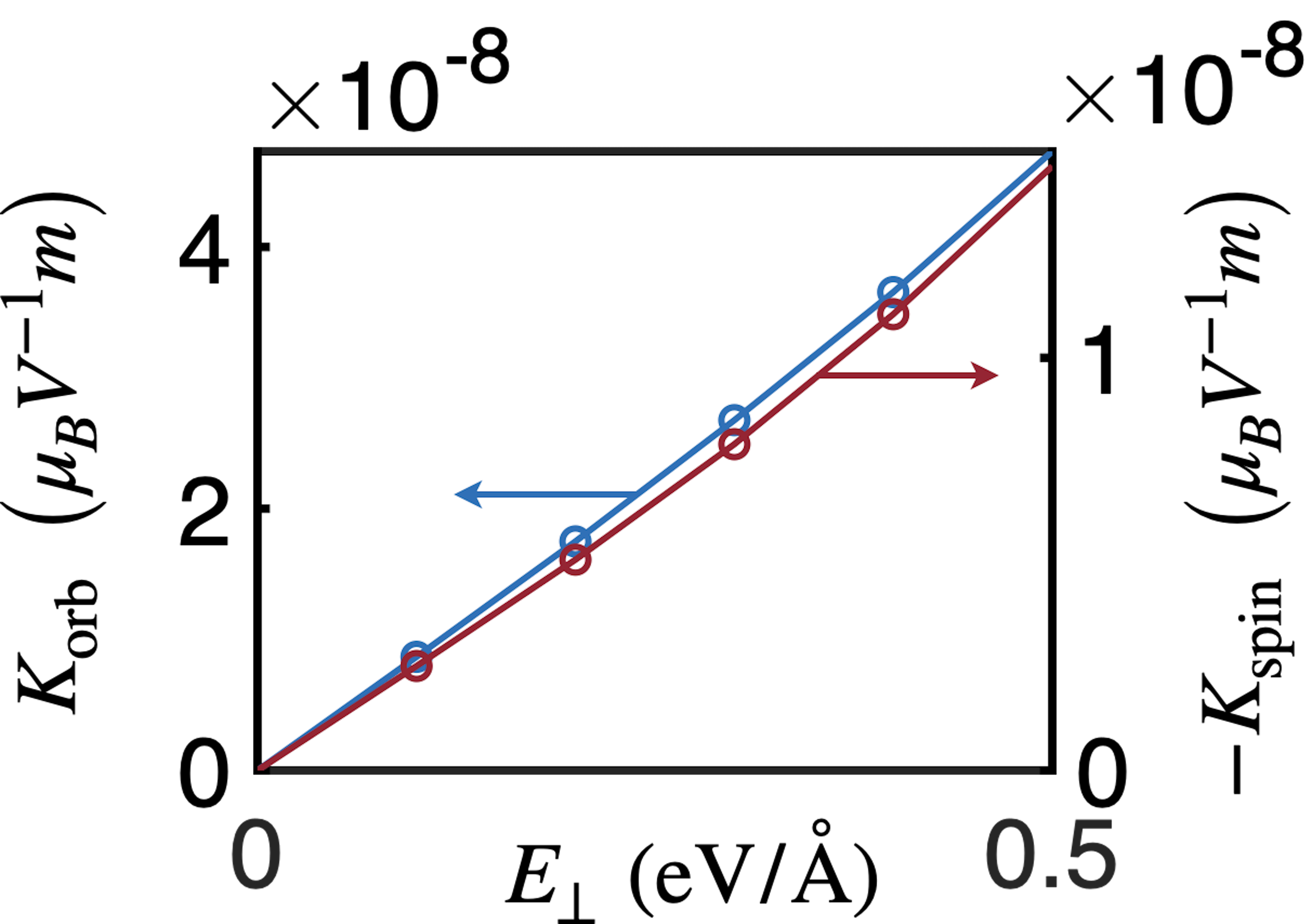} 
   \caption{Variation of the orbital and spin Edelstein response with the gate field $E_\perp$ for MoS$_2$ for electron doping with $N_e = 0.02$ electrons per unit cell.
    } 
    \label{fig:EE-field dependence}
\end{figure}

\begin{figure*}[tb]
    \centering
    \includegraphics[width=0.6\linewidth]{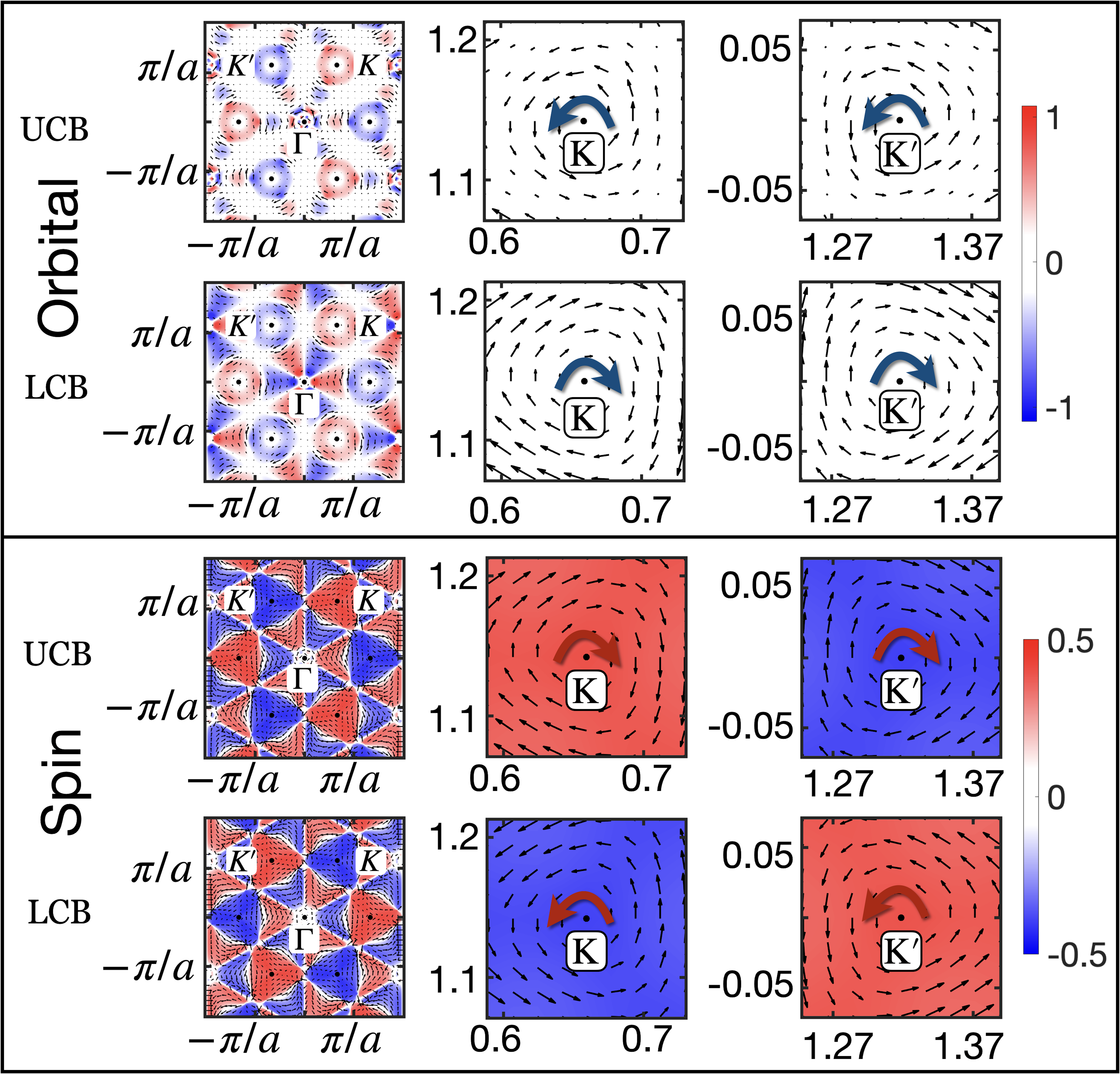}
    \caption{Orbital/spin texture for the lower  and upper  conduction band (LCB/ UCB) for MoS$_2$. The conduction band minimum is at the ${\rm K/K'}$ point, and the two bands
    have slightly different energies  due to a small spin splitting (see  Fig. \ref{schematic-2}). }
    \label{fig:cond_texture}
\end{figure*}

It has been demonstrated\cite{doping01} that it is also possible to n-dope the TMDs, although the doping concentration is less than  that for the holes. The maximum demonstrated p-doping is $\approx 4 \times 10^{14}/ cm^2$, while the same for n-doping is $\approx 2 \times 10^{13}$/ cm$^2$, the latter corresponding to doped electron concentration of $N_e \approx 0.018$ electrons per unit cell area.
One notable feature of the electron doped case is that the OEE is several times stronger than the SEE, as seen from Fig. \ref{fig:EE-field dependence}.

For the electron doped system, the relevant valley is  ${\rm K/K'}$, since it forms the bottom of the conduction bands. There is a dip in energy in the conduction bands along the ${\rm K}-\Gamma$ line, but it only gets occupied beyond  a certain dopant concentration, and the $\Gamma$ conduction band that lies way above in energy does not come into play for experimentally feasible doping concentrations. Thus our results in this section are valid for small electron concentrations ($n_e \lesssim 0.02$), since our downfolded Hamiltonian model Eq.(\ref{eq:hfull}) does not reproduce the ${\rm K}-\Gamma$ dip.
For higher dopant concentrations, this energy dip needs to be properly described. For this,
higher neighbor interactions need to be retained in the tight-binding theory and the corresponding downfolded Hamiltonian.

Fig. \ref{fig:cond_texture} shows the spin/orbital texture for the conduction bands obtained
from the Hamiltonian (\ref{eq:hfull}), and the chiralities at the 
valence band bottom at ${\rm K/K'}$ are highlighted. 
Unlike the hole-doped case, in the present case, perturbation theory does not work too well due to the proximity of the perturbing states close above the conduction minimum.
We have instead obtained the spin/orbital textures close to ${\rm K/K'}$  from the numerical solution of the full Hamiltonian Eq. (\ref{eq:hfull}). 
For small momentum and small gate fields ($\gamma_1 \propto E_\perp$), the
spin chirality is opposete for the two bands, $\nu = \pm 1$,
 \begin{equation}
    \langle \vec S_\parallel \rangle_{\rm K, K'} =  -c \  \nu \gamma_1 q \hat \theta,
        \label{eq:S-electron}
 \end{equation}
 where $c > 0$ is a material-dependent proportionality constant.
 The opposite spin chirality produces a cancellation effect between the two bands, just like for the hole-doped case, resulting in a diminished 
 Edelstein effect in the spin channel.
 The orbital channel shows a similarly opposite chirality between the two bands. However, unlike the spin case Eq. (\ref{eq:S-electron}), the orbital moment, which can be expressed as
  \begin{equation}
    \langle \vec L_\parallel \rangle_{\rm K, K'} =  c_\nu (q) \nu \gamma_1  \hat \theta,
    \label{eq:L-electron}
 \end{equation} 
 has a different prefactor $c_\nu (q)$ for the two bands $\nu = \pm 1$,
which  does not produce a similar cancellation effect.

The corresponding Edelstein susceptibilities, in units of
$    e \tau \hbar^{-1} A_0$ and includig both bands,
are given by
\begin{eqnarray}
    K_{\rm orb} &=& c_o \gamma_1 N_e,  \nonumber \\
         K_{\rm spin} &=& c_s \gamma_1 (N_e^+ - N_e^-),
 \label{eq:K-electron}
\end{eqnarray}
where $N_e^\nu$ is the dopant density (electrons per unit cell area) in the individual bands, 
total electron concentration $N_e = N_e^+ + N_e^- $, and $c_o$ and $c_s$ are material dependent constants. 
As seen from Eq. (\ref{eq:K-electron}), the spin response   from the two bands has a cancellation effect, and $K_{\rm spin}$ is non-zero only because the two spin-split bands have a slightly different dopant concentration due to the small energy separation between them.

The computed Edelstein susceptibility for MoS$_2$ is shown in Fig. (\ref{fig:summary})
for both electron and hole doping. For the electron doping case, $K_{\rm orb}$ increases linearly with the dopant concentration $N_e$, while $K_{\rm spin}$ remains relatively constant
and much smaller in magnitude due to the cancellation effect seen from Eq. (\ref{eq:K-electron}). 
Unlike the hole case, the magnitude of the OEE is relatively large even for small doping and it is also several times larger than the corresponding SEE.
Furthermore, the linear dependence of the $K_{\rm orb}$ with $N_e$ suggests a way to differentiate between OEE and SEE in the experiments.

For the hole-doped case, the $\Gamma$ valley has the dominant contribution as Fig. (\ref{fig:summary}) shows.
As already mentioned, the $\Gamma$ valley can be moved up in energy with a small compressive strain, in which case both $K_{\rm orb}$ and $K_{\rm spin}$
are significantly enhanced, even for a small hole doping as the holes occupy the $\Gamma$ valley.
Thus, for the hole doped case, a small compressive strain will enhance the OEE by a significant amount, while for the electron doped case, a strain is not needed for a large OEE. 
In Table \ref{Table3}, we compare the computed strengths of the Edelstein susceptibility for several materials with a fixed dopant concentration. 
The results indicate certain materials such as MoTe$_2$ to have larger OEE as compared to others.

We note that the OEE in the TMDs, both for electron and hole doping, is much stronger than what has been reported for actual material systems in the literature, which makes the TMDs ideal candidates to study the OEE.
To our knowledge, concrete calculations exist for two material systems: (i) The two-dimensional electron gas at the SrTiO$_3$ interface\cite{oee4} ($K_{\rm orb} \sim 10^{-9} \ \mu_B\,\text{V}^{-1}\text{m}$) and (ii) A bilayer system with Rashba spin–orbit interaction\cite{oee8} ($K_{\rm orb} \sim  10^{-11} \ \mu_B\,\text{V}^{-1}\text{m}$).
In contrast, for the present materials, we find that the orbital Edelstein susceptibility can be higher
by one or two order magnitude, $K_{\rm orb}$ typically $\sim 10^{-8}- 10^{-7} \ \mu_B\,\text{V}^{-1}\text{m}$ as seen from Fig. \ref{fig:summary}. 
In many of our calculations, we have employed the gate field $E_\perp = 0.4$ eV/\AA, which is quite high but has been demonstrated to be possible for the TMDs\cite{gate01}, they being two-dimensional materials.
However, since the Edelstein response is roughly proportional to the gate field, a gate field even a factor of 10 smaller than the maximum possible, would still lead to a large OEE response. 
This suggests the TMDs to be excellent materials for studying the OEE as well as the SEE.

\begin{table}
 \caption{Edelstein susceptibility, $K_{\rm orb}$ and $K_{\rm spin}$,  in units of $ 10^{-8} \mu_B V^{-1}m$, for the electron-doped materials. Results are given for  the doping concentration $N_e  = 0.02$ electrons per unit cell area. Parameter are: the gate field  $E_\perp = 0.2$ eV/\AA\ and  $\tau = 12$ ps. 
 }
    \centering
    \begin{tabular}{c | c c c c c c}
        \hline 
         Material & MoS$_2$ & MoSe$_2$  & MoTe$_2$  &   WS$_2$  & WSe$_2$  &  WTe$_2$   \\ 
         \hline
         $K_{\rm orb}$  &  {1.31}  & {4.48} & {8.12} & {0.31}  & {2.15}  & {5.76}  \\
        $K_{\rm spin}$  &  {-0.38} & {-2.04}  & {-5.77} & {-0.08}  & {-1.39} & {-4.14} \\
         \hline
    \end{tabular}
    \label{Table3}
\end{table}



\section{Summary and Conclusion}

\begin{figure*}[tbh]
    \centering
\includegraphics[scale=0.22]{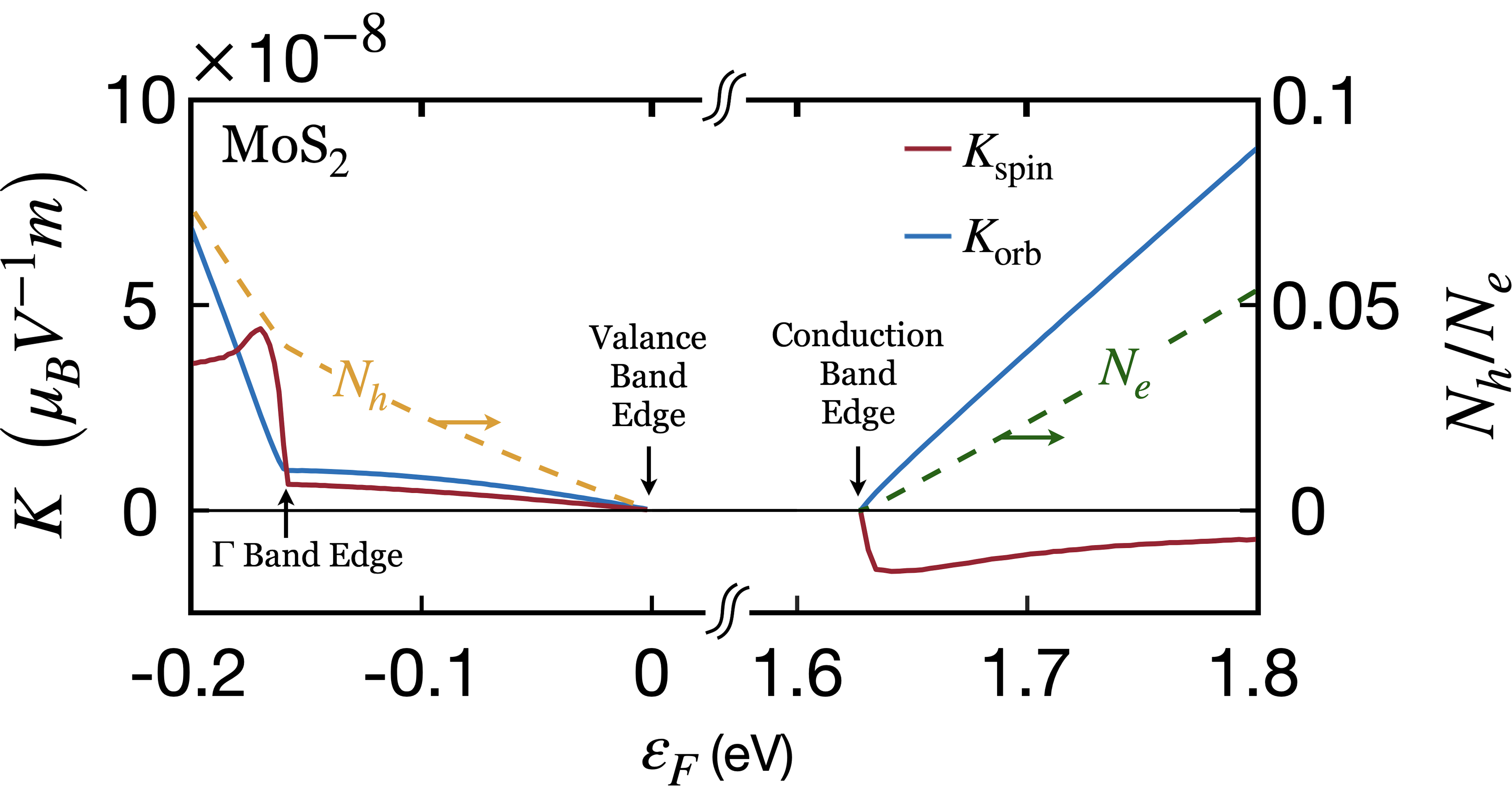}
    \caption{Edelstein susceptibilities, $K_{\rm orb}$  and $K_{\rm spin}$ for MoS$_2$,  as a function of $\varepsilon_F$ for both electron and hole doped case,
    The colored dashed lines indicate the dopant concentration $N_e / N_h$ (electrons or holes per unit cell area of the crystal) corresponding to the Fermi energy $\varepsilon_F$.  
Parameters were:  gate field $E_\perp = 0.4$ eV/\AA\ and relaxation time $\tau = 16$ ps. Both 
  $K_{\rm orb}$  and $K_{\rm spin}$ are proportional to $E_\perp$ for small fields; So the results can be scaled for other values of $E_\perp$. }
    \label{fig:summary}
\end{figure*}

In this paper, we studied the orbital and spin Edelstein effects in doped monolayer TMDs under a gate field normal to the monolayer. Both electron and hole doping were considered. 
The gate-field induced broken $\sigma_h$ symmetry  is also broken due to a substrate and is also naturally present in the Janus materials\cite{Pratik}.
The broken symmetry
results in a momentum dependent chiral spin/orbital texture  with moments parallel to the plane.  This in turn leads to the spin and orbital Edelstein effects, caused by a non-equilibrium shift of the Fermi surface in response to an in-plane charge current. 
The normal component $L_z/S_z$ does not produce an Edelstein effect, unless there is a uniaxial strain\cite{Bhowal-GME}.

We employed a minimal $5 \times 5$ Hamiltonian model within the metal $d$ orbital subspace. The form of the Hamiltonian  in the presence of the gate field are given by Eqs. (\ref{eq:h1}-\ref{matrixelements2}), and the Hamiltonian parameters for a number of compounds obtained by fitting with the density-functional bands are listed in Table \ref{Table1}. 
A simpler form Eq. (\ref{eq:Ising-Rashba}) is valid at the band edges in the gap region, which is a combination of an Ising and a Rashba term.
The full $d^5$ manifold is needed to be retained as the gate field couples the  $L = 1$ angular momentum sector with the $L=2$ and $0$ sector. 



We found a chiral spin/orbital texture with a moment component parallel to the plane, the strength of which is proportional to the gate field. 
For both electron and hole doping, a robust Edelstein effect in both the spin and orbital channel was obtained with the Edelstein susceptibilities, $K_{\rm orb}$, $K_{\rm spin}$ as large as $\sim 10^{-8}- 10^{-7} \ \mu_B\,\text{V}^{-1}\text{m}$, which is much larger than what has been reported in the literature for other materials. 
This, together with the fact that it is possible  to achieve substantial  electron and hole doping as well as a large gate field up to $0.4$ eV/\AA, suggests the TMDs to be an ideal platform to study the orbital Edelstein effect. 

If the doping concentration is small, then for unstrained sample, electron doping is better than hole doping in producing a larger Edelstein effect (as summarized in Fig. \ref{fig:summary}). 
This is due to the fact that the conduction bands at ${\rm K/K'}$ that the doped electrons 
occupy are affected much more strongly by the electric field (controlled by the parameter $\gamma_1$), while the ${\rm K/K'}$ hole pockets are affected much less, being controlled by $\gamma_2 \ll \gamma_1$.
On the other hand, it is possible to achieve a much larger $p$ doping (0.35 per unit cell area for holes vs. 0.02 for electrons\cite{doping01}), so that for the high $p$ doping case, the $\Gamma$ pocket becomes occupied, which has a larger Edelstein response.
So, for hole doping and unstrained structure, a hole concentration larger than the critical value $N_c^h$ listed in Table \ref{Table2} is desirable.  This critical value $N_c^h$ needed to access the $\Gamma$ bands is for some materials way too high and unphysical, while for others it is experimentally accessible.

However,  strain has an important advantage for the hole doped case in amplifying the Edelstein effect, and a small amount of strain can be utilized to enhance the effect, even when the dopant concentration is small.
We showed that a small compressive strain quickly raises the energy of the $\Gamma$ valence bands, reducing the $\Gamma-{\rm K}$ energy separation and making the $\Gamma$ bands accessible even for a small hole concentration. For instance, for MoS$_2$ (see Table {\ref{Table2}}), as little as  one percent of compressive strain makes the $\Gamma$ valley the top of the valence bands, causing the doped holes to go there and thus enhancing the Edelstein response substantially. In contrast, for the electron doped case,  strain is not expected to make any significant difference as the $\Gamma$ valley is too high in energy.
Apart from the  fundamental significance, our results are   relevant for spintronics applications on system involving the TMD materialss\cite{Mudgal, Shao, Zhang}.


\section{Acknowledgments} We thank Dr. Pratik Sahu for helpful discussions. This work was initiated while one of us (SS) was a visitor at the Indian Institute of Technology-Madras, India, under  the VAJRA fellowship scheme of the Science and Engineering Research Board, Government of India.



\begin{thebibliography}{10}
\bibitem{Note-error} Eqs. (\ref{matrixelements1}) and 
(\ref{matrixelements2}) correct some typographical errors that were present in Ref. \cite{Shanavas}.
\end{thebibliography}
\end{document}